\documentclass[preprint2]{aastex}

\shorttitle{Limits of Adaptive Optics for high contrast imaging}
\shortauthors{O. Guyon}
\usepackage{graphicx}
\usepackage[mathscr]{eucal}
\begin{document}

\title{Limits of Adaptive Optics for high contrast imaging}
       
\author{Olivier Guyon}

\affil{Subaru Telescope, 650 N. A'ohoku Place, Hilo, 96720 HI, USA}
\email{guyon@naoj.org}

\begin{abstract}
The effects of photon noise, aliasing, wavefront chromaticity and scintillation on the point spread function (PSF) contrast achievable with ground based adaptive optics (AO) are evaluated for different wavefront sensing schemes.
I show that a wavefront sensor (WFS) based upon the Zernike phase contrast technique offers the best sensitivity to photon noise at all spatial frequencies, while the Shack-Hartmann WFS is significantly less sensitive.
In AO systems performing wavefront sensing in the visible and scientific imaging in the near-IR, the PSF contrast limit is set by the scintillation chromaticity induced by Fresnel propagation through the atmosphere. On a 8m telescope, the PSF contrast is then limited to 1e-4 to 1e-5 in the central arcsecond. Wavefront sensing and scientific imaging should therefore be done at the same wavelength, in which case, on bright sources, PSF contrasts between 1e-6 and 1e-7 can be achieved within 1 arcsecond on a 8m telescope in optical/near-IR. The impact of atmospheric turbulence parameters (seeing, wind speed, turbulence profile) on the PSF contrast is quantified. I show that a focal plane wavefront sensing scheme offers unique advantages, and I discuss how to implement it. Coronagraphic options are also briefly discussed. 
\end{abstract}

\section{Introduction}
High contrast imaging of the immediate environment (within a few astronomical units) of nearby stars is critical to the understanding of formation and evolution of planetary systems. The ultimate goal of planetary systems studies is to find and characterize planets similar to ours, in the hope that we can find other ``habitable'' words susceptible of harboring life. Several approaches are currently under development to achieve the high level of contrast required :
\begin{itemize}
\item{Nulling interferometry in the mid-IR with a ~30m baseline space interferometer.}
\item{Visible coronagraphy with a 4m to 8m space telescope.}
\item{Large ground based telescopes (8m to 100m) and high performance AO systems optimized for bright targets.}
\end{itemize}
While the first 2 options are targeting Earth-size planets around nearby stars, the ground based systems have more modest initial goals: planets more massive than Jupiter or young Jupiter mass planets. The plans to build larger (30m to 100m) telescopes and the fast progress in high-performance AO systems and coronagraphy however opens up the possibility of pursuing more ambitious goals. Direct imaging of Earth-size planets is even considered for 100m diameter telescopes \citep{gilm04,hawa03}.

The contrast detection limit within a PSF is set by photon noise and speckle noise in the image. If only photon noise is considered the theoretical detection limit for a large telescope (30m to 100m) allows relatively easy detection of Earth-size planets around nearby stars \citep{ange03,hawa03}. This however seems to be a very optimistic assumption, as current AO systems are all limited by speckle noise \citep{raci99} within the central arcsecond. The detection limit is therefore likely to be driven by how well the speckles can be calibrated/removed. Fast atmospheric speckles average fairly rapidly, but slower-evolving speckles are more problematic. Experience acquired on ground-based telescope has shown that there is no such thing as a truly static speckles, as extremely small drifts in the wavefront are sufficient to appreciably change the speckle's intensity: for a speckle of intensity $10^{-5}$ the central star's intensity to be stable to within 1\%, the corresponding spatial frequency would need to be stable to $2.5\:10^{-6}$ wave (4 picometers at 1.6 $\mu$m).
  
In non-coronagraphic imaging with current AO systems on large telescopes, the PSF wings are relatively smooth in long exposures: about 10\% of the speckle background does not average at about 1\arcsec (figure 1 in \cite{bocc03}). Through careful PSF calibration, some of the residual speckle structure can be subtracted \citep{roth01}, yielding a point source detection limit more than 10  times fainter than the PSF background level within the central arcsecond. This factor tends to become larger with increasing distance from the PSF core, partially thanks to the chromatic elongation of speckles which makes the PSF smoother. 
Differential imaging techniques based on spectral properties of the source could further increase this factor by up to $10^2$ for simultaneous imaging in 2 bands and $10^4$ for simultaneous imaging in 3 bands \citep{maro00}. Techniques using the coherence properties of speckles have also been proposed \citep{bocc98,guyo04}.

In this study, performance of an adaptive optics system is quantified by the ratio between the light intensity at the point of the PSF considered and the light intensity at the PSF's center. This quantity is referred to as the PSF contrast in the rest of the paper.
The goal of this work is to give limits on this PSF contrast achievable with AO systems, and to propose solutions to reach these limits: which WFS to choose ? how to drive the deformable mirror (DM) ? is a coronagraph necessary ? if yes, which one ? Detection limits for faint companions are significantly harder to predict than PSF contrast for the reasons detailed above, and will not be computed in this paper. 

The PSF contrast in a photon-noise limited AO system is a function of the ability of the WFS to accurately measure the corresponding spatial frequency in the pupil plane phase. Analytical expressions of the fundamental contrast limits imposed by photon noise and chromaticity of the wavefront are derived is \S\ref{sec:contrastanalytical}. Aliasing effects are discussed in \S\ref{sec:alias}, and solutions to reduce their impact on the PSF contrast are proposed. In \S\ref{sec:WFSs}, the sensitivity of common WFSs to photon noise is discussed, and a WFS based upon Zernike's phase contrast is shown to offer optimal sensitivity. 
Results of \S\ref{sec:contrastanalytical}, \S\ref{sec:alias} and \S\ref{sec:WFSs} are combined and discussed in \S\ref{sec:perf_disc} to derive realistic limits to the PSF contrast in ground-based AO systems and identify optimal approaches to detect extrasolar planets. \S\ref{sec:perf_disc} shows that a focal plane WFS can be especially advantageous, and this option is discussed in more detail in \S\ref{sec:fpwfs_layout}. In \S\ref{sec:coronagraphy}, I discuss the need for coronagraphy and the choice of the correct coronagraph to reach the PSF contrast derived in this study.

\section{Fundamental limits of wavefront sensing for high contrast AO: analytical expressions}
\label{sec:contrastanalytical}
\subsection{Speckles and wavefronts}
Notations used in this work are given in table \ref{tab:notations}. In this paper, I consider a ``perfect'' coronagraph: if the entrance pupil of the system had no phase aberrations, the focal plane light intensity would be equal to 0 outside the Inner Working Angle (smallest angular separation at which the coronagraph can be used to detect a faint companion, denoted IWA in the rest of the paper) of the coronagraph within the spectral bandwidth considered. All observations are made at the zenith: atmospheric dispersion is not taken into account.

The pupil plane complex amplitude is denoted
\begin{equation}
\label{equ:ppca}
W(\vec{u}) = \mathscr{A}(\vec{u}) \: e^{i \phi(\vec{u})}
\end{equation}
where $\mathscr{A}(\vec{u})$ is the amplitude and $\phi(\vec{u})$ the phase of the wavefront.
The pupil plane phase aberration
\begin{equation}
\label{equ:phiuv}
\phi(\vec{u}) = \frac{2 \pi h}{\lambda} \cos\left(2 \pi \vec{f}\vec{u}+\theta\right)
\end{equation}
creates 2 symmetric images of the central PSF \citep{malb95}:
\begin{equation}
\label{equ:syms}
I(\vec{\alpha}) = PSF(\vec{\alpha}) + \left(\frac{\pi h}{\lambda}\right)^2 [PSF(\vec{\alpha}+\vec{f}\lambda)+PSF(\vec{\alpha}-\vec{f}\lambda)]
\end{equation}
where $\lambda$ is the imaging wavelength, $h \ll \lambda$ is the amplitude (in meter) of the sine wave phase aberration of spatial frequency $\vec{f}$, and $\vec{\alpha} = \vec{f}\lambda$ is the angular coordinate on the sky. The phase of these speckles are $\pi/2-\theta$ and $\pi/2+\theta$.
Similarly, a multiplicative amplitude error 
\begin{equation}
\label{equ:ampuv}
\mathscr{A}(\vec{u}) = 1 + a \cos\left(2 \pi \vec{f}\vec{u}+\theta\right)
\end{equation}
creates 2 symmetric speckles of phases $-\theta$ and $\theta$ on either side of the ``ideal'' image $PSF(\vec{\alpha})$:
\begin{equation}
\label{equ:syms1}
I(\vec{\alpha}) = PSF(\vec{\alpha}) + \left(\frac{a}{2}\right)^2 [PSF(\vec{\alpha}+\vec{f}\lambda)+PSF(\vec{\alpha}-\vec{f}\lambda)].
\end{equation}
To simplify notations in the rest of the paper, I will denote $f=|\vec{f}|$, $u=|\vec{u}|$ and $\alpha=|\vec{\alpha}|$.

If a perfect DM is used, the ability of the AO system to suppress light at the position $\vec{\alpha}$ is then given by the ability of the WFS to measure the phase and amplitude of the corresponding spatial frequency in the pupil plane phase. In an optically ``perfect'' system (no wavefront or amplitude errors introduced by the optical elements, noiseless detector), the effects that limit the performance of WFSs for high contrast imaging are:
\begin{itemize}
\item{{\bf Photon Noise in the WFS.}}
\item{{\bf Chromaticity} of the optical pathlength difference (OPD) and amplitude between the WFS wavelength $\lambda$ and the imaging wavelength $\lambda_i$.}
\item{{\bf Aliasing.} The wavefront measurement is corrupted by higher spatial frequency aberrations that propagate into modes which are detected by the WFS.}
\end{itemize}
In this section, the first two effects are discussed, while aliasing effects are studied separately in \S\ref{sec:alias}. Table \ref{tab:contrast_contrib} lists the terms computed analytically in this section:
\begin{itemize}
\item{$C_0$: PSF contrast limit imposed by OPD aberrations in uncorrected atmospheric turbulence.}
\item{$C_1$: PSF contrast limit imposed by amplitude aberrations in uncorrected atmospheric turbulence (scintillation).}
\item{$C_2$: PSF contrast limit imposed by residual OPD aberrations after AO correction. This term is computed analytically in this section as a function of the WFS sensitivity to photon noise $\beta_p$. Using the technique proposed in appendix \ref{app:wfsw}, $\beta_p$ is computed for different WFSs in \S\ref{sec:WFSs}.}
\item{$C_3$: PSF contrast limit imposed by residual amplitude aberrations after AO correction of OPD and amplitude.}
\item{$C_4$: PSF contrast limit imposed by the differential OPD between the WFS and imaging wavelengths. This term is caused by the chromaticity of Fresnel propagation.}
\item{$C_5$: PSF contrast limit imposed by the differential scintillation between the WFS and imaging wavelengths. This term is caused by the chromaticity of Fresnel propagation.}
\item{$C_6$: PSF contrast limit imposed by the differential OPD between the WFS and imaging wavelengths. This term is caused by the chromaticity of the refraction index of air.}
\end{itemize}
The terms $C_i$ are computed for high levels of correction (Strehl ratio $\approx$ 1), which allows simplification of most equations. The final PSF contrast $C$, computed as a function of angular separation, is then obtained as follows:
\begin{itemize}
\item{{\bf No AO correction:} $C=C_0+C_1$.}
\item{{\bf AO correction of phase only:} $C=C_1+C_2+C_4+C_6$.}
\item{{\bf AO correction of phase and amplitude:} $C=C_2+C_3+C_4+C_5+C_6$.}
\end{itemize}

\begin{deluxetable}{lll}
\tabletypesize{\small}
\tablecaption{\label{tab:notations} Notations and units}
\tablewidth{0pt}
\startdata
\hline
& symbol & unit \\
\hline
Telescope diameter & $D$ & m\\
Detection wavelength & $\lambda_i$ & m\\
WFS wavelength & $\lambda$ & m\\
Seeing wavelength & $\lambda_0$ & m\\
Fried parameter & $r_0$ at $\lambda_0$ & m\\ 
Wind speed & $v$ & m.s$^{-1}$\\
Source brightness & $F$ & ph.s$^{-1}$.m$^{-2}$\\
Angular separation & $\alpha$ & rad\\
Turbulence altitude & $z$ & m\\
Turbulence profile & $C_n^2(z)$ & \\
WFS sensitivity to OPD & $\beta_p$ & \\
WFS sensitivity to amplitude & $\beta_a$ & \\
\enddata
\end{deluxetable}

\begin{deluxetable}{ll}
\tabletypesize{\small}
\tablecaption{\label{tab:contrast_contrib} Contributions to the PSF contrast}
\tablewidth{0pt}
\startdata
\hline
Uncorrected atmospheric OPD & $C_0$ \\
Uncorrected atmospheric amplitude & $C_1$ \\
Residual atmospheric OPD after correction & $C_2$ \\
Residual atmospheric amplitude after correction & $C_3$\\
OPD chromaticity (Fresnel propagation) & $C_4$\\ 
Scintillation chromaticity (Fresnel propagation) & $C_5$\\ 
Refraction index chromaticity & $C_6$\\ 
\enddata
\end{deluxetable}

\subsection{Uncorrected atmospheric turbulence ($C_0$ and $C_1$)}
In the paraxial approximation, Fresnel propagation of wave of complex amplitude $W(\vec{u},0)$ over a distance $z$ produces a wave $W(\vec{u},z)$ described by
\begin{equation}
\label{equ:fresnelprop}
W(\vec{u},z) = W(\vec{u},0) \otimes \exp(i \pi u^2 / z \lambda)
\end{equation}
where $\otimes$ is the convolution operator. This is equivalent to a phase shift of each spatial frequency component of the wavefront by
\begin{equation}
d\phi = \pi \:f^2\:z\:\lambda.
\end{equation}
A pupil plane complex amplitude
\begin{equation}
W(\vec{u},0) = 1 + i \frac{2 \pi h}{\lambda} \: \sin\left(2\pi \vec{u}\vec{f} + \theta \right)
\end{equation}
therefore becomes
\begin{eqnarray}
\label{equ:fresnel1}
W(\vec{u},z) = 1+\sin(d\phi)\times\frac{2 \pi h}{\lambda}\sin\left(2\pi \vec{u}\vec{f} + \theta \right) 
\nonumber \\+ i \cos(d\phi) \times \frac{2 \pi h}{\lambda}\sin\left(2\pi \vec{u}\vec{f} + \theta \right).
\end{eqnarray}
This equation shows that Fresnel propagation of a pure sine wave phase aberration produces both an amplitude and a phase aberration of identical spatial frequency in the pupil plane. This effect is periodic for each spatial frequency, as $W(\vec{u},z+z_T) = W(\vec{u},z)$ with $z_T = 2/f^2\lambda$ the Talbot distance \citep{talb36}.

For ground-layer Komogorov atmospheric turbulence, the power spectrum of the 2D phase is
\begin{equation}
\phi(f) = \frac{0.023}{{r_0}^{5/3}} \: f^{-11/3}
\end{equation}
where $r_0$ is the Fried parameter. In a telescope pupil of diameter $D$, the power given by a single spatial frequency is obtained by integration of $\phi(f)$ over a 2D domain of width proportional to $1/D$. Through numerical simulations, the corresponding amplitude (in meter) of the sine-wave component of spatial frequency $f$ is computed:
\begin{equation}
\label{equ:A}
h(f) = \frac{0.22 \: \lambda_0}{f^{11/6} \: D \: {r_0}^{5/6}}.
\end{equation}
where $\lambda_0$ is the wavelength at which $r_0$ is measured.

Taking into account the Fresnel propagation given in equation \ref{equ:fresnel1}, the following expressions are obtained for the OPD and amplitude components of atmospheric turbulence in equations \ref{equ:phiuv} and \ref{equ:ampuv}:
\begin{equation}
\label{equ:h}
h(f) = \frac{0.22 \: \lambda_0}{f^{11/6} \: D \: {r_0}^{5/6}} \sqrt{X(f,\lambda_i)}
\end{equation}
\begin{equation}
\label{equ:a}
a(f) = \frac{2\pi \: 0.22 \: \lambda_0}{\lambda_i \: f^{11/6} \: D \: {r_0}^{5/6}} \sqrt{Y(f,\lambda_i)}
\end{equation}
where
\begin{equation}
X(f,\lambda_i) = \frac{\int C_n^2(z) \cos^2(\pi z f^2 \lambda_i) dz}{\int C_n^2(z) dz}
\end{equation}
and
\begin{equation}
Y(f,\lambda_i) = \frac{\int C_n^2(z) \sin^2(\pi z f^2 \lambda_i) dz}{\int C_n^2(z) dz} = 1-X(f,\lambda_i).
\end{equation}
Since Fresnel diffraction is chromatic, $X$ and $Y$ are function of $\lambda_i$. $X$ is the fraction of the atmospheric turbulence which produces phase errors, the remaining part producing amplitude errors (scintillation). For low altitude turbulence and/or low spatial frequencies, $X \approx 1$: the beam propagation length is too short to allow Fresnel diffraction to transform phase errors in amplitude errors.
By combining equations \ref{equ:syms} and \ref{equ:h}, since $f=\alpha/\lambda_i$, atmospheric phase aberrations produce the following contrast at $\lambda_i$:
\begin{equation}
C_0(\alpha) = \frac{0.484 \: \pi^2 \: {\lambda_0}^2 \: {\lambda_i}^{5/3} \: X(\alpha/\lambda_i,\lambda_i)}{\alpha^{11/3}\:D^2\:{r_0}^{5/3}}.
\end{equation}
Similarly, from equations \ref{equ:syms1} and \ref{equ:a}, the amplitude aberrations (scintillation) produce the following contrast at $\lambda_i$:
\begin{equation}
C_1(\alpha) = \frac{0.484 \: \pi^2 \: {\lambda_0}^2 \: {\lambda_i}^{5/3} \: Y(\alpha/\lambda_i,\lambda_i)}{\alpha^{11/3}\:D^2\:{r_0}^{5/3}}.
\end{equation}
Since $X+Y=1$, the combined contribution of phase and amplitude aberrations in the PSF contrast is independent of the turbulence altitude.

\subsection{Effect of WFS photon noise and time lag on the corrected phase ($C_2$)}
\label{sec:C2}
In the Taylor approximation used in this work, atmospheric turbulence is moving in front of the telescope pupil at a speed $v$ (wind speed along the direction $\vec{\alpha}$ considered). 
In a closed-loop AO system, the corrected amplitude $h_c$ of the spatial frequency considered is quadratic sum of a component due to time lag and a component due to photon noise (given by equation. \ref{equ:WFS_quality}) :
\begin{equation}
\label{equ:hc}
h_c = \sqrt{\left(2 \pi \: h(f) \: v \: t \: f\right)^2 + \left(\frac{\lambda}{2 \: \pi}\right)^2 \left(\frac{\beta_p}{\sqrt{t \: F \: \pi \: D^2/4}}\right)^2}
\end{equation}
where $t$ is the WFS sampling time, $F$ is the source brightness (in $ph.s^{-1}.m^{-2}$) and $D$ is the telescope diameter.

$h_c$ is minimal for
\begin{equation}
t_h = \left(\frac{\lambda}{\lambda_0}\right)^{2/3}\frac{0.204 \: \beta_p^{2/3} \: {r_0}^{5/9} \: {f}^{5/9}}{F^{1/3}\:v^{2/3} \: X(\alpha/\lambda_i,\lambda_i)^{1/3}}
\end{equation}

The corresponding residual error produces a symmetric pair of speckles (equation \ref{equ:syms}) with a contrast to the central PSF peak:
\begin{equation}
\label{equ:C2}
C_2(\alpha) = 0.7475 \: \frac{\lambda^{4/3}\:{\lambda_0}^{2/3}\:\beta_p^{4/3}\:v^{2/3}\:X(\alpha/\lambda_i,\lambda_i)^{1/3}}{{\lambda_i}^{13/9}\:F^{2/3}\:D^2\:{r_0}^{5/9}\:\alpha^{5/9}}.
\end{equation}
where $\lambda_i$ is the wavelength at which the final image is obtained, and might be different from $\lambda$, the wavefront sensing wavelength.


\subsection{Effect of WFS photon noise and time lag on the corrected amplitude ($C_3$)}
The light intensity distribution (scintillation) in the pupil plane can be measured and corrected for. In the frozen turbulence flow model adopted in this work, scintillation is an amplitude screen moving in front of the telescope. The effect of photon noise and time lag on the corrected amplitude $a_c$ can therefore be written
\begin{equation}
\label{equ:ac}
a_c = \sqrt{(2\pi\:a(f)\:v\:t\:f)^2+\frac{\beta_a^2}{t\:F \: \pi \: D^2/4}}.
\end{equation}
This equation is identical to equation \ref{equ:hc} if $h_c$ and $h(f)$ are replaced by $\lambda_i a_c /2\pi$ and $\lambda_i a(f)/2\pi$ respectively. The optimal sampling time is therefore
\begin{equation}
\label{equ:t0_a}
t_a = \left(\frac{\lambda}{\lambda_0}\right)^{2/3}\frac{0.204 \: \beta_a^{2/3} \: {r_0}^{5/9} \: {f}^{5/9}}{F^{1/3}\:v^{2/3} \: Y(\alpha/\lambda_i,\lambda_i)^{1/3}},
\end{equation}
and the corresponding contrast $C_3$ is
\begin{equation}
\label{equ:C3}
C_3(\alpha) = 0.7475 \: \frac{\lambda^{4/3}\:{\lambda_0}^{2/3}\:\beta_a^{4/3}\:v^{2/3}\:Y(\alpha/\lambda_i,\lambda_i)^{1/3}}{{\lambda_i}^{13/9}\:F^{2/3}\:D^2\:\alpha^{5/9}\:{r_0}^{5/9}}.
\end{equation}

\subsection{Chromaticity of OPD and scintillation}

\subsubsection{OPD chromaticity produced by Fresnel propagation ($C_4$)}
Fresnel propagation is chromatic, and the OPD at the telescope pupil is therefore chromatic. When perfectly corrected at one wavelength (the WFS wavelength), the OPD in the imaging wavelength will show a small residual which limits the achievable contrast to:
\begin{equation}
C_4(\alpha) = \frac{C_0(\alpha) \: dX(\alpha/\lambda_i,\lambda_i,\lambda)}{X(\alpha/\lambda_i,\lambda_i)} 
\end{equation}
where
\begin{equation}
dX(f,\lambda_i,\lambda) = \frac{\int C_n^2(z) \left(\cos(\pi z f^2 \lambda_i)-\cos(\pi z f^2 \lambda)\right)^2 dz}{\int C_n^2(z) dz}.
\end{equation}

\subsubsection{Scintillation chromaticity ($C_5$)}
Similarly, Fresnel propagation produces wavelength-dependent intensity variations in the pupil plane. This produces a limit $C_5$ on the achievable contrast :
\begin{equation}
\label{equ:C5}
C_5(\alpha) = \frac{C_1(\alpha) \: dY(\alpha/\lambda_i,\lambda_i,\lambda)}{Y(\alpha/\lambda_i,\lambda_i)}.
\end{equation}
where 
\begin{equation}
dY(f,\lambda_i,\lambda) = \frac{\int C_n^2(z) \left(\sin(\pi z f^2 \lambda_i)-\sin(\pi z f^2 \lambda)\right)^2 dz}{\int C_n^2(z) dz}
\end{equation}

\subsubsection{Chromaticity of the air refraction index ($C_6$)}
The index of refraction of dry air at standard temperature and pressure is wavelength-dependent \citep{edle66}:
\begin{equation}
n(\lambda) = 1.0 + 8.34213\:10^{-5} + \frac{0.0240603}{130-\lambda^{-2}} + \frac{0.00015997}{38.9-\lambda^{-2}}.
\end{equation}
The corresponding PSF contrast is
\begin{equation}
C_6(\alpha) =  C_0(\alpha) \left(\frac{n(\lambda_i)-n(\lambda)}{1-n(\lambda_i)}\right)^2
\end{equation}

\section{Aliasing effects}
\label{sec:alias}
\subsection{WFS aliasing}
The pupil OPD and amplitude aberrations can only be corrected by the AO system below a cutoff spatial frequency $f_c$, because of limited sampling in the pupil plane DM and/or in the WFS. For the contrast expressions derived in \S\ref{sec:contrastanalytical} to be applicable, the measurement accuracy of a pupil plane phase aberration of spatial frequency $f<f_c$ at the WFS wavelength $\lambda$ must be limited by photon noise. Unfortunately, measurement of an OPD or amplitude aberration of frequency $f<f_c$, even in the absence of photon noise, can be corrupted by aliasing: other spatial frequencies (usually above $f_c$, but not always) can create a WFS signal at frequency $f$.

The optical part of the WFS (before the detector) does not produce aliasing: a phase aberration at frequency $f$ only creates an optical signal of frequency $f$ in the pupil plane. This signal can be spot displacements (for Shack-Hartmann WFS) or intensity modulation (for curvature WFS for example). Aliasing is therefore produced by the limited sampling of the measurement in the pupil plane.

Two approaches exist to suppress or mitigate aliasing in WFSs:
\begin{itemize}
\item{{\bf Increasing the WFS spatial sampling.}
If a good detector (low readout noise and dark current) is used, this solution can be highly successful for all but one of the WFS considered in this work. The single exception is the Shack-Hartmann WFS, where an increase of spatial sampling in the pupil plane (smaller subapertures) increases the measurement error on low-order modes due to photon-noise (see section \S\ref{sec:WFS_SH}).}
\item{{\bf Preventing spatial frequencies above $f_c$ to be ``seen'' by the optical part of the WFS.} This can be done by spatial filtering in the focal plane \citep{poyn04} or by using an anti-aliasing optical filter before the detector \citep{taka05}. These solutions reduce aliasing on all pupil-plane WFSs, and are most effective if the WFS sampling is regular, as is usually the case. Curvature WFSs include a focal plane iris at the vibrating membrane (usually to reduce stray light and sky background), which can be used to reduce aliasing in high-Strehl regime. Anti-aliasing optical filters are routinely used in imaging with CCDs, and are often placed immediately before the detector in commercial digital cameras. Anti-aliasing optical filters with total rejection of high spatial frequencies can be designed \citep{lege97} for monochromatic light, and their performance in white light is still good: the solution proposed by \citet{lege97} reduces aliasing by a factor 24 with a 20\% bandpass.}
\end{itemize}

\subsection{Algorithm used to compute aliasing-free DM control signals}
\label{sec:DMcontrol}
\begin{figure}[htb]
\includegraphics[scale=0.24]{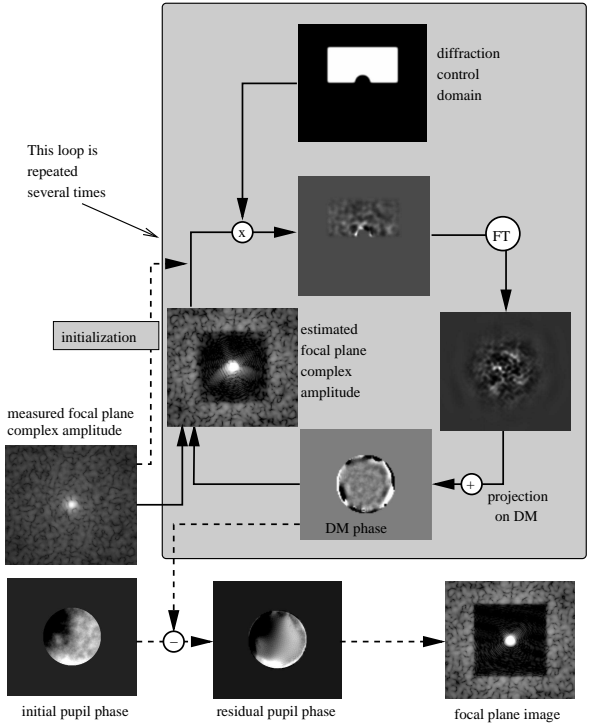}
\caption{\label{fig:driveDM_algo} Proposed algorithm used to drive the DM in a closed-loop AO system.}
\end{figure}

If the focal plane complex amplitude (or, equivalently, the aliasing-free pupil plane complex amplitude) is perfectly known up to a spatial frequency $f_c$, it is possible to drive a DM to cancel focal plane speckles within a region of the image corresponding to spatial frequencies lower than $f_c$. \citet{malb95} proposed to use a non-linear minimization algorithm to control the DM. In this work, a Gershberg-Saxton algorithm \citep{ger72} is proposed to find the optimal DM control signals. Since this method is tailored at finding the solution of a problem with constraints on both a function (pupil plane complex amplitude) and its Fourier transform (focal plane complex amplitude), it seems naturally well adapted \citep{fauc89}. The steps of the algorithm are shown in figure \ref{fig:driveDM_algo}:
\begin{itemize}
\item{(1) The region of the focal plane within which the speckles are to be canceled is first chosen. This ``diffraction control domain'' (DCD) should exclude the central part of the PSF and should not extend beyond the pupil spatial frequency defined by the DM actuator size or WFS sampling. This region can be within a half plane if amplitude errors (scintillation) in the pupil plane are expected, or can include both sides of the PSF for correction of OPD aberrations only.}
\item{(2) The focal plane complex amplitude is multiplied by the DCD to represent the complex amplitude that should be canceled by the DM.}
\item{(3) The 2D complex function computed in step (2) is Fourier transformed to produce the ``ideal'' pupil plane complex amplitude required to cancel the speckles within the DCD.}
\item{(4) The pupil plane complex amplitude function is ``projected'' on the DM: for each actuator of the DM, the phase which best matches the function computed in step (3) is used to update the DM state.}
\item{(5) Using the updated DM state and the initial measured focal plane complex amplitude, the updated complex amplitude in the focal plane is numerically estimated, and steps (2) to (5) can then be repeated.}
\end{itemize}

This algorithm converges very rapidly: only a few iterations are required to obtain a high contrast (about $10^{-10}$) if the initial PSF aberrations are low. It is also computationally less greedy than the solution proposed by \citet{malb95} (the computing time is dominated by 2 Fourier transforms), and can therefore be implemented in a fast closed loop control system.

\begin{figure}[htb]
\includegraphics[scale=0.22]{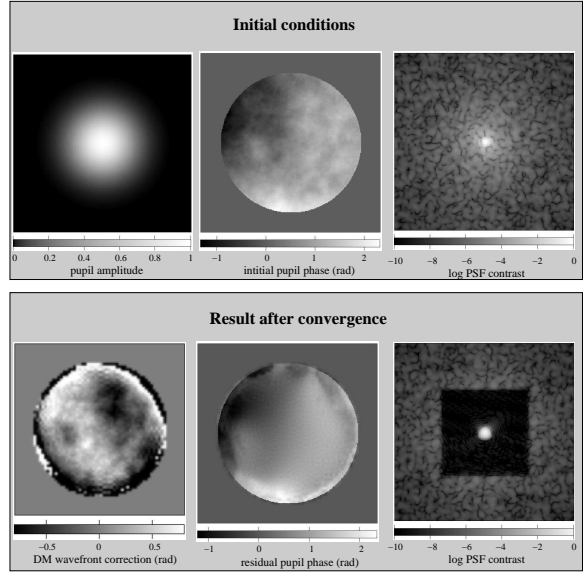}
\caption{\label{fig:DMdrive_perf} Example of diffracted light suppression using the algorithm detailed in \S\ref{sec:DMcontrol}. In this case, the DM stroke is limited to $\pm 0.4$ radian ($\pm 0.8$ radian of phase correction), and the initial pupil phase aberration is about 3 radian from peak to peak. The DM actuators are square-shapes, and their influence functions are overlapping (each DM influence function is a square convolved by a Gaussian). In this example, there are 50 actuators across the diameter of the pupil (2000 actuators total).}
\end{figure}

This algorithm is also very flexible and performs well in non-ideal conditions. Non-ideal DM characteristics such as irregular actuator shapes, ``dead'' actuators or coupled influence functions can easily be included in the algorithm (step (4)) with minimal cost in complexity or computation time. For example, figure \ref{fig:DMdrive_perf} shows that a solution yielding good PSF contrast can be found even if the PSF has large aberrations and if the DM has insufficient stroke to fully correct them. This particular example illustrates how the algorithm is able to find a solution for which the diffraction within the DCD is canceled even though the residual phase aberrations in the pupil plane are still large: these aberration are confined to either high or low spatial frequencies, but are very small in the spatial frequency range corresponding to the DCD. It should be noted that in such non-ideal conditions, the number of iterations (steps (2) to (5)) required to converge can be quite high (about 100 in this example).

\subsection{Closed loop operation}
In a closed loop AO system, the DM control algorithm proposed in \S\ref{sec:DMcontrol} is used to compute frequent but small updates of the DM: the phase function in the pupil plane needed to cancel the speckles is very small. The number of iterations (steps (2) to (5) of the algorithm) required within the algorithm is therefore small (a single iteration is sufficient). The computing time can consequently be made compatible with kHz update rate on modern computers for systems with up to $~10^4$ actuators: on a 128x128 actuators system (13000 actuators on the circular pupil) with a focal plane image sampling such that the DCD occupies 256x256 pixels, the time required for the 2 Fast Fourier Transforms is about 1ms on a modern computer.

The number of photons per focal plane speckle is typically less than 10 per sampling period, and the corresponding relative error on the measured complex amplitude of the speckle due to photon noise is more than 10\%. The closed loop performance of the AO correction is therefore not sensitive to small errors introduced by the algorithm described in \S\ref{sec:DMcontrol}. Even a 5\% error in the knowledge of the DM response has a negligible impact on the system performance.

\section{Wavefront sensors sensitivities}
\label{sec:WFSs}
The sensitivity of a WFS is a quantitative measure of how photon noise affects its measurement of OPD or amplitude. In this section, the sensitivity $\beta_p$ of WFSs for OPD sensing (when only OPD is measured by the WFS) is computed. $\beta_p$ is used in equation \ref{equ:C2} to estimate the contribution $C_2$ of WFS photon noise to the PSF contrast. For each WFS, I show how $C_2$ varies as a function of angular separation, and how WFS design parameters affect it. An exact definition of $\beta_p$ and details on how it is computed are given in appendix \ref{app:wfsw}. The sensitivity $\beta_a$ for amplitude sensing is given within the discussion in \S\ref{sec:wfs_beta_disc}.

The results obtained in this section are only valid for small residual phase variance at the wavefront sensing wavelength.

\subsection{Shack-Hartmann WFS}
\label{sec:WFS_SH}
In a Shack-Hartmann WFS (SHWFS), the quantities $I_k$ measured are spot displacements. The associated noises for a diffraction-limited spot, in the absence of background light, are \citep{hard98}
\begin{equation}
\sigma_{I_k} = \frac{0.277 \: \lambda}{d_{sa} \: \sqrt{N_{sa}}}
\end{equation} 
for a continuous noiseless detector and
\begin{equation}
\sigma_{I_k} = \frac{0.500 \: \lambda}{d_{sa} \: \sqrt{N_{sa}}}
\end{equation} 
for a quad cell detector. In the above equations, $d_{sa}$ is the subaperture size and $N_{sa}$ is the number of photons per subaperture. To account for atmospheric turbulence within each cell, $1/d_{sa}$ should be replaced by $\sqrt{1/{d_{sa}}^2+1/{r_0}^2}$. 

Using the equations detailed in appendix \ref{app:wfsw}, the following results are obtained:
\begin{itemize}
\item{For a Shack-Hartmann WFS with a continuous noiseless detector:
\begin{equation}
\label{equ:SHbeta1}
\beta_{p} = \frac{0.67}{f\:d_{sa}}\:\sqrt{1+\left(\frac{d_{sa}}{r_0}\right)^2}
\end{equation} 
}
\item{For a Shack-Hartmann WFS with noiseless quad cells:
\begin{equation}
\beta_{p} = \frac{1.48}{f\:d_{sa}}\:\sqrt{1+\left(\frac{d_{sa}}{r_0}\right)^2}
\end{equation} 
}
\end{itemize}
where $d_{sa}$ is the subaperture size. For both equations, $f\:d_{sa} \leq 1/3$ (minimum of 3 lenslet per sine wave period), as lower pupil plane sampling increase $\beta_{p}$. For example, with 2 lenslets per period, if the center of lenslets coincide with the crests and peaks of the sine wave phase aberration, no signal will be produced by the SHWFS.

With a SHWFS, photometry of the spots can be used to measure amplitude variations in the pupil plane, without altering the accuracy of the phase measurement: the sensitivity $\beta_p$ is maintained even if OPD and scintillation are measured simultaneously.

\begin{figure}[htb]
\hspace{-0.15in}\includegraphics[scale=0.32]{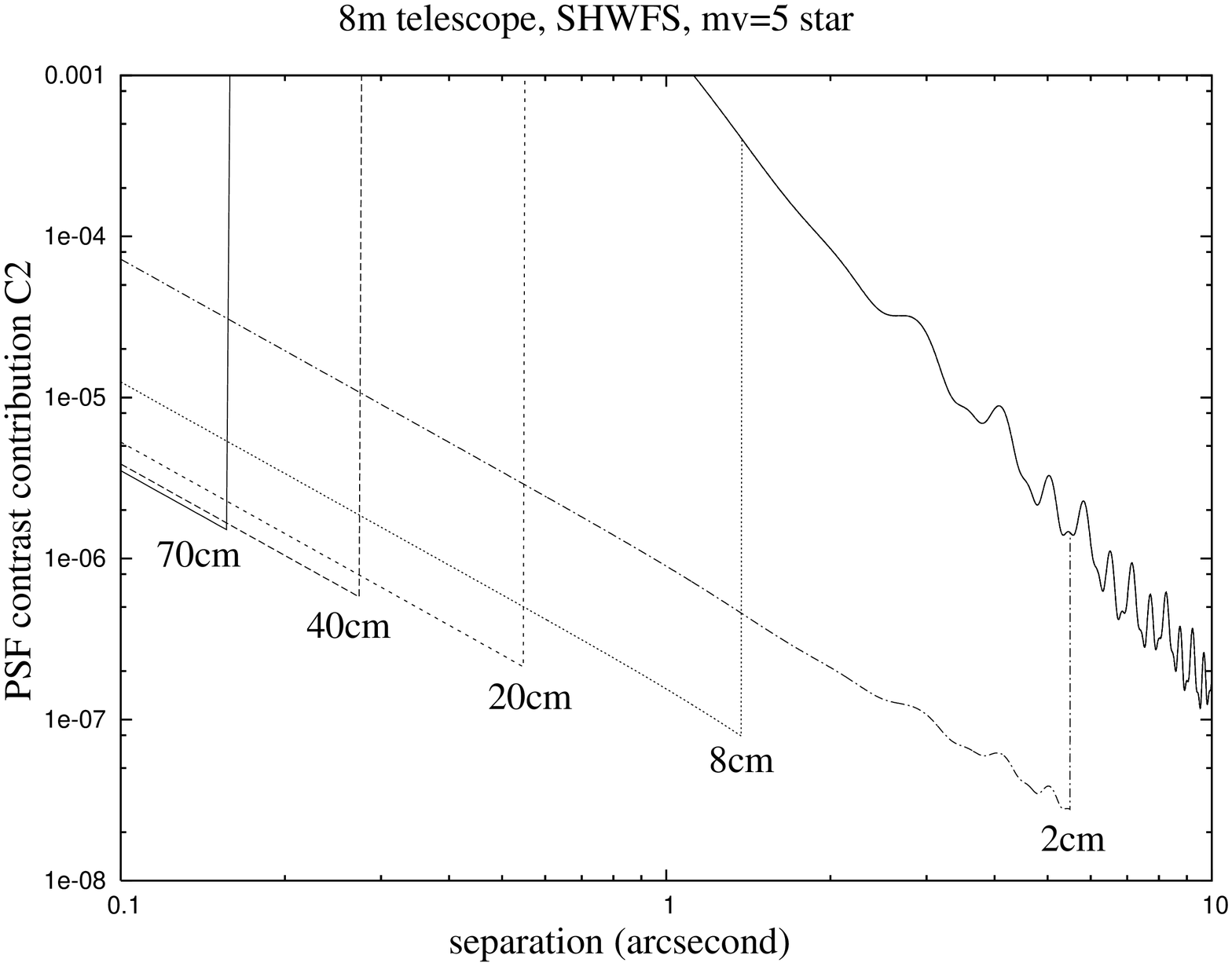}
\caption{\label{fig:contrast_sh8m} PSF contrast component $C_2$ with Shack-Hartmann WFS using various subaperture sizes. The continuous line shows the PSF contrast component $C_0$ without correction of the atmospheric turbulence. Parameters used for this simulation are listed in table \ref{tab:simul_param}.}
\end{figure}

The PSF contrast component $C_2$ achievable with a SHWFS is shown in figure \ref{fig:contrast_sh8m} for subaperture sizes ranging from 2cm to 70cm. for each subaperture, a continuous noiseless detector was assumed, rather than a less sensitive quad-cell. 
In the inner region of the PSF, the contrast $C_2$ decreases as the -17/9 power of angular separation (equations \ref{equ:SHbeta1} and \ref{equ:C2}). No correction is possible beyond the sampling limit of the WFS: the contrast $C_2$ reaches a minimum value at this transition point. The constrast $C_2$ at small angular separations is independant of the number of subapertures if the subaperture size is larger than the seeing. However, if subapertures are smaller than $r_0$, diffraction by each subaperture increases the subaperture's focal plane spot size and therefore reduces the sensitivity of the WFS. It therefore seems impossible to simultaneously optimize the contrast over a wide range of separations.

To achieve the optimal performance shown in figure \ref{fig:contrast_sh8m}, the wavefront integration time $t_h$ needs to be proportionnal to $\alpha^{-1/9}$.

\subsection{Curvature wavefront sensor (CWFS)}
\label{sec:curv}

In a curvature WFS \citep{rodd88,rodd91}, a spherical phase aberration is introduced in the focal plane, which is equivalent, in the pupil plane, to Fresnel propagation. The pupil image is therefore ``conjugated'' to an altitude which is set by the amplitude of the focal plane phase aberration. Equation \ref{equ:fresnel1} shows that Fresnel propagation of a pure sine wave phase aberration produces both an amplitude and a phase aberration of identical spatial frequency in the pupil plane. The curvature WFS therefore transforms phase aberrations into light intensity modulations in the pupil plane.

The WFS measures intensities $I_k$ in the pupil plane. I assume here that $N$ such measurements are taken per spatial period:
\begin{equation}
I_k = \frac{N_{ph}}{N} \left(1+\frac{4\:\pi\:A\:\sin(d\phi)}{\lambda} \sin\left(\frac{2\pi k}{N}+\phi\right)\right)
\end{equation}
with $\sigma_{I_k} = \sqrt{N_{ph}/N}$.

Using the method detailed in appendix \ref{app:wfsw}, the following expression for $\beta_p$ is obtained:
\begin{equation}
\label{equ:SigmaCVp}
\beta_{p}(\alpha) = \sin^{-1}\left(\frac{\pi\:\delta z\:\lambda \: \alpha^2}{{\lambda_i}^2}\right)
\end{equation}
where $\delta z$ is the conjugation altitude of the pupil plane (in curvature AO systems, 2 pupil plane images are usually acquired, at conjugation altitudes $+\delta z$ and $-\delta z$). This result is independant of $N$ for $N>2$. 

\begin{figure}[htb]
\hspace{-0.0in}\includegraphics[scale=0.32]{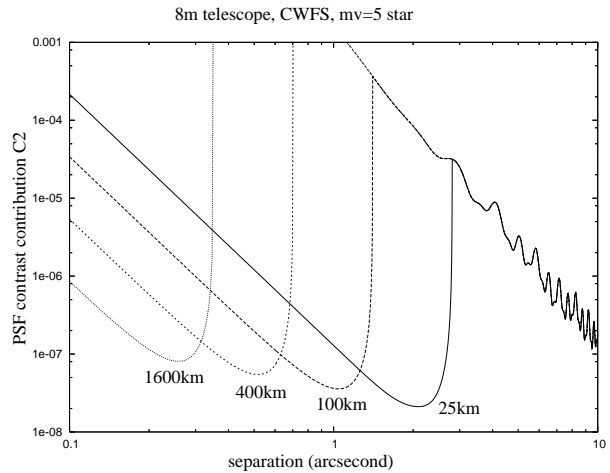}
\caption{\label{fig:contrast_cv8m} PSF contrast component $C_2$ obtained with a curvature WFS using defocalization distances ranging from 25km to 1600km. Parameters used for this simulation are listed in table \ref{tab:simul_param}.}
\end{figure}

The PSF contrast component $C_2$ achievable with a CWFS is shown in figure \ref{fig:contrast_cv8m}. The amount of defocus introduced in the focal plane of the CWFS can be adjusted to tune its sensitivity to an optimal spatial frequency in the pupil plane (for which the term in the sine of equation \ref{equ:SigmaCVp} is $\pi/2$).
In the inner regions of the PSF, the contrast $C_2$ decreases as the -29/9 power of the angular separation (equations \ref{equ:SigmaCVp} and \ref{equ:C2}), which is significantly steeper than for a SHWFS. This steep increase of wavefront error at low spatial frequencies is also referred to as ``noise propagation'', and is known to be more serious for CWFS than for SHWFS. Soon after the contrast reaches a minimum, the defocus distance becomes too large (the sine in equation \ref{equ:SigmaCVp} becomes close to zero) and no reliable correction can be applied to the wavefront. Theoretically, correction of higher spatial frequencies is possible as the sine in equation \ref{equ:SigmaCVp} periodically oscillates between 1 and -1, but this possibility was not considered in figure \ref{fig:contrast_cv8m}: in a real CWFS, spectral bandwidth and time evolution of $\delta z$ (usually closer to a sine wave than a step function) prevent this feature from being usable.

At small angular separations, the wavefront integration time $t_h$ for a CWFS needs to be proportionnal to $\alpha^{-7/9}$ to achieve the result shown in figure \ref{fig:contrast_cv8m}. This suggests that a curvature WFS can greatly benefit from a modal control scheme, where the correction speed can be adjusted for each spatial frequency, as opposed to a zonal reconstruction with a fixed integration time.

Although the optimal constrast region is narrower in a CWFS than it is with a SHWFS, is also is deeper: at a given separation, a properly tuned CWFS performs better than a SHWFS. This is especially true close to the PSF center, where a ``tuned'' CWFS can reach a sensitivity $\beta_p=1$. Since changing the extrafocal distance in a CWFS is usually very easy, it is in fact possible to continuously move the optimal contrast region between small and large angular separations during an observation. The equivalent technique would be optically more complex in a SHWFS, as the subapertures size would need to be modified.

\subsection{Pyramid WFS}
\label{sec:pyramid}
The pyramid WFS \citep{raga96} divides the focal plane in 4 quadrants, each one beeing then reimaged in a separate pupil plane. In the geometrical optics approximation, wavefront slopes can be measured as contrast between pairs of pupil images. The focal plane point which defines the position of the quadrants (the ``center'' of the pyramid) can be rapidly rotating around the PSF core to increase linearity and dynamical range, at the expense of sensitivity.

\begin{figure}[htb]
\hspace{-0.0in}\includegraphics[scale=0.4]{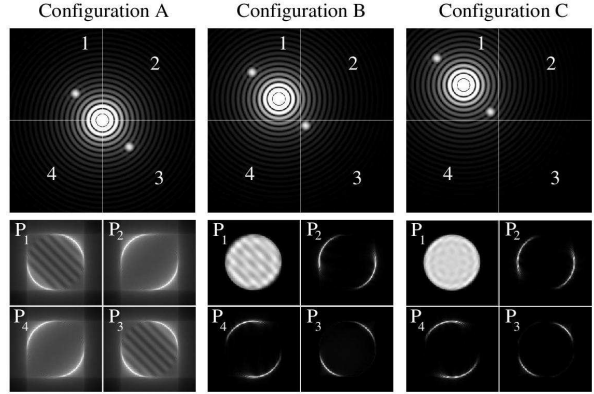}
\caption{\label{fig:pyram} Focal plane images (top) and corresponding pupil images $P_i$ (bottom) for a sine-wave pupil phase error (corresponding to 2 symmetric speckles in the focal plane). See text for details.}
\end{figure}

I denote $P_i(x,y)$ the pupil image corresponding to quadrant $i$, as shown in figure \ref{fig:pyram}, and $P_{ref}(x,y)$ the pupil image in the absence of a focal plane pyramid. The pyramid WFS can be operated in 2 ways:
\begin{itemize}
\item{{\bf Fixed pyramid position.} The top of the pyramid (junction point between the 4 quadrants) is at the center of the PSF core, corresponding to configuration A in figure \ref{fig:pyram}.}
\item{{\bf Modulation of pyramid position.} The top of the pyramid is moving on a circle of radius $r_p$. Configurations B (the central PSF core and one speckle are within the same quadrant) or C (the quadrant containing the PSF core either contains no speckle or both symmetric speckles) can then occur.}
\end{itemize}
\subsubsection{Fixed pyramid position}
I denote $P_i^0$ the pupil image corresponding to quadrant $i$ obtained in configuration A in the absence of phase aberrations for a unobstructed circular pupil. $P_i^0$ consists of a fainter pupil image with bright sharp edges and a significant fraction of the light diffracted outside the geometric pupil. Although the total light in each pupil image $P_i^0$ is one quarter of the original pupil image, the total light within the geometric pupil (excluding the bright edges) is about 6\% of the original pupil. I consider a pupil complex amplitude 
\begin{equation}
W(\vec{u},0) = 1 + i \frac{2 \pi A}{\lambda} \: \sin\left(2\pi \vec{u}\vec{f} + \phi \right)
\end{equation}
corresponding to a set of 2 symmetric speckles as shown in figure \ref{fig:pyram}. In configuration A, these 2 speckles interfere with $P_i^0$ in quadrants 1 and 3: 
\begin{equation}
P_1 = P_1^0 + \sqrt{P_1^0 \: P_{ref}} \times \frac{2 \pi A}{\lambda} \: \sin\left(2\pi \vec{u}\vec{f} + \phi \right)
\end{equation}
\begin{equation}
P_2 = P_2^0
\end{equation}
\begin{equation}
P_3 = P_3^0 - \sqrt{P_3^0 \: P_{ref}} \times \frac{2 \pi A}{\lambda} \: \sin\left(2\pi \vec{u}\vec{f} + \phi \right)
\end{equation}
\begin{equation}
P_4 = P_4^0
\end{equation}

The pyramid WFS therefore directly measures the pupil phase \citep{veri04,veri05}: this is somewhat different from the geometrical optics understanding of this concept, in which phase slope is measured by the pupil images. Due to the splitting of the focal plane into 4 zones, reconstruction of the full wavefront map requires $P_1$, $P_2$, $P_3$ and $P_4$. For example, pupil images $P_1$ and $P_3$ are only sensitive to pupil phase spatial frequencies corresponding to zones 1 and 3 of the pyramid.

Using the method detailed in appendix \ref{app:wfsw},
\begin{equation}
\label{equ:SigmaPY}
\beta_{p} = \sqrt{2}.
\end{equation}

\subsubsection{Modulation of pyramid position}
\begin{figure}[htb]
\hspace{-0.0in}\includegraphics[scale=0.4]{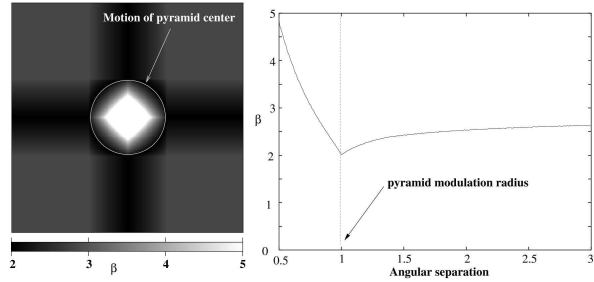}
\caption{\label{fig:betapyram} Value of $\beta_p$ in a modulated pyramid WFS. A 2D map of $\beta_p$ is shown on the left, and a averaged radial profile is plotted on the right.}
\end{figure}

I consider a motion of the pyramid center on a circle of radius $r_p > \lambda/d$, with no change in the orientation of the pyramid.
As shown in figure \ref{fig:pyram}, configurations B and C occur as the pyramid moves. In configuration B on figure \ref{fig:pyram}, within the geometric pupil:
\begin{equation}
P_1 = P_{ref} \times \left( 1 + \frac{2 \pi A}{\lambda} \: \sin\left(2\pi \vec{u}\vec{f} + \phi \right) \right)
\end{equation}
\begin{equation}
P_2 = P_3 = P_4 = 0
\end{equation}
In configuration C, $P_i = P_{ref}$ for the pyramid zone containing the PSF core, and $P_i = 0$ for the other pupil images.
In a long exposure (longer than the modulation time of the pyramid position):
\begin{equation}
P_i = P_{ref} \left(f_i^B \left( 1 \pm \frac{2 \pi A}{\lambda} \: \sin\left(2\pi \vec{u}\vec{f} + \phi \right) \right) + f_i^C \right)
\end{equation}
where $f_i^B$ and $f_i^C$ are the fraction of the time during which the PSF core is in the pyramid quadrant $i$ and the configuration is B and C respectively. The sign of the modulated intensity signal is opposite between $P_1$ and $P_3$, and between $P_2$ and $P_4$. Since $f_i^B+f_i^C = 0.25$ (the PSF core spends a quarter of its time on each zone of the pyramid), in a long exposure (longer than the modulation time of the pyramid position):
\begin{equation}
P_i = \frac{P_{ref}}{4} \left( 1 \pm 4 f_i^B \frac{2 \pi A}{\lambda} \: \sin\left(2\pi \vec{u}\vec{f} + \phi \right) \right).
\end{equation}
Each of the 4 pupil images contains the intensity modulation, but with different signal levels. 
Since $f_1^B = f_3^B$ and $f_2^B = f_4^B$, using the method detailed in appendix \ref{app:wfsw},
\begin{equation}
\label{equ:SigmaPYm}
\beta_p = \frac{2 \sqrt{2}}{\sqrt{(4f_1^B)^2+(4f_2^B)^2}}.
\end{equation}
Figure \ref{fig:betapyram} shows how $\beta_p$ varies across the focal plane. It is minimum at the pyramid modulation radius and increases rapidly toward the center of the PSF.

\subsubsection{Discussion}
\begin{figure}[htb]
\hspace{-0.15in}\includegraphics[scale=0.32]{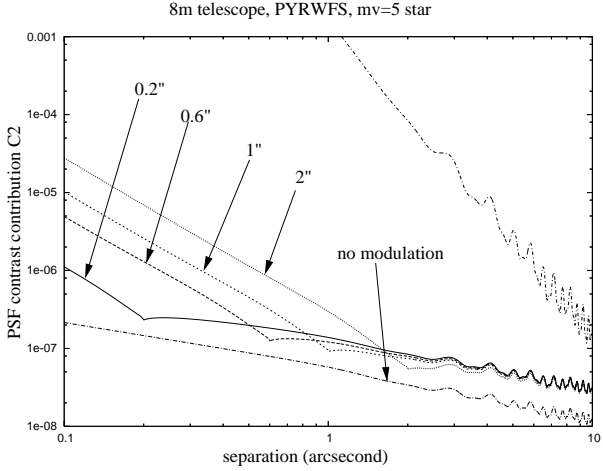}
\caption{\label{fig:C2pyr8m} PSF contrast component $C_2$ obtained with a pyramid WFS with and without modulation of the pyramid's position. $C_2$ is plotted for modulation radii ranging from 0\farcs2 to 2\arcsec. The top curve shows the PSF contrast $C_0$ corresponding to uncorrected turbulence phase aberration. Parameters used for this simulation are listed in table \ref{tab:simul_param}.}
\end{figure}

Figure \ref{fig:C2pyr8m} shows the contrast component $C_2$ for modulated and fixed pyramid WFSs. The sensitivity of the pyramid WFS is better if the pyramid is fixed, and this mode of operation should be preferred in high-contrast AO on bright sources, as the linearity range of the WFS is then not a concern. If the pyramid is fixed, it may be replaced by a ``roof top'' (a pyramid with only 2 faces), which would offer the same sensitivity with 2 pupil images instead of 4.

\subsection{Mach-Zehnder pupil plane interferometer}
In this wavefront sensing scheme suggested by \citet{ange94}, a beam splitter produces two copies of the same wavefront. One of the copies is spatially filtered and interferometrically combined with the unfiltered wavefront. Wavefront phase is transformed into intensity variations in the 2 pupil images produced by the interferometer. Minimum sensitivity is reached when the first beam splitter is symmetric (50/50 beam splitter), and the interferometer's OPD is such that the 2 pupil images have the same brightness when the wavefront is perfect. The interferometer's OPD may be modulated to increase dynamical range.

The WFS measures intensities $I_k$ in the two pupil plane. I assume here that $2N$ such measurements are taken per spatial period ($N$ measurements per pupil image):
\begin{equation}
I_k = \frac{N_{ph}}{2 N} \left(1 \pm \sin\left(\frac{2\:\pi\:A}{\lambda}\right) \sin\left(\frac{2\pi k}{N}+\phi\right)\right)
\end{equation}
with $\sigma_{I_k} = \sqrt{N_{ph}/2N}$, and the sign in front of the sine is different for each pupil image.

Using the method detailed in appendix \ref{app:wfsw}, the following expression for $\beta_p$ is obtained:
\begin{equation}
\label{equ:SigmaMZ}
\beta_p = 2.
\end{equation}


Figure \ref{fig:C2phase8m} shows the PSF contrast component $C_2$ obtained with a PPMZWFS.

\subsection{Focal plane WFS}
\label{sec:WFS_FP}
In a FPWFS, the amplitude and phase of focal plane speckles are measured by inducing interferences between the focal plane complex amplitude and a set of known ``reference waves'' \citep{ange03}. Optical configurations to produce the reference waves and measure the interferences are discussed in \S\ref{sec:fpwfs_layout}.

The amplitude and phase of a focal plane speckle created by the sine-wave pupil phase error are
\begin{equation}
A_s = \sqrt{N_{ph}}\frac{\pi h}{\lambda}
\end{equation}
\begin{equation}
\phi_s = \phi
\end{equation}
to which correspond the real and imaginary parts of the speckle
\begin{equation}
x_s = A_s \cos(\phi_s) = \sqrt{N_{ph}} \: \frac{x_0}{2}
\end{equation}
\begin{equation}
y_s = A_s \sin(\phi_s) = \sqrt{N_{ph}} \: \frac{y_0}{2}
\end{equation}
where $x_0$ and $y_0$ follow the notations used in appendix \ref{app:wfsw}. $x_0$ and $y_0$ are estimated through the measurement of $N$ intensities:
\begin{equation}
I_k = (\frac{x_s}{\sqrt{N}}+x_k)^2 + (\frac{y_s}{\sqrt{N}}+y_k)^2
\end{equation}
where $k = 0...N-1$, and $x_k,y_k$ are the $N$ reference waves with which the speckle light interferes. The total number of photons in the speckle is equally shared between the $N$ measurements: $x_s$ and $y_s$ are therefore divided by $\sqrt{N}$.

\begin{equation}
\frac{d I_k}{d x_0} = \frac{x_k \: \sqrt{N_{ph}}}{\sqrt{N}}
\end{equation}
\begin{equation}
\frac{d I_k}{d y_0} = \frac{y_k \: \sqrt{N_{ph}}}{\sqrt{N}}
\end{equation}

Since the measurement noise on $I_k$ is $\sqrt{{x_k}^2+{y_k}^2}$ (photon noise), the estimate of $x_s$ and $y_s$ from the measurement of $I_k$ is insensitive to the amplitude of the reference wave (but not its phase). For example, multiplying the light level in the reference wave by 4 will double $\frac{d I_k}{d x_s}$ and $\frac{d I_k}{d y_s}$, and will also double the measurement noise $\sqrt{{x_k}^2+{y_k}^2}$. The problem is therefore reduced to finding a set of phases for the reference waves. 

Numerical simulations using the method detailed in appendix \ref{app:wfsw} show that a minimum of 2 reference waves are needed. The optimal performance (minimum value of $\Sigma$) is reached when the 2 waves are offset by $\pi/2$:
\begin{equation}
\beta_p = 2.
\end{equation}
Increasing the number of waves does not lead to better solutions. If the wavefront does not contain amplitude variations (no scintillation), the symmetry property of the focal plane speckles allows the use of 2 speckles to sense a single pupil plane spatial frequency, in which case:
\begin{equation}
\beta_p = \sqrt{2}.
\end{equation}

\begin{figure}[htb]
\hspace{-0.15in}\includegraphics[scale=0.32]{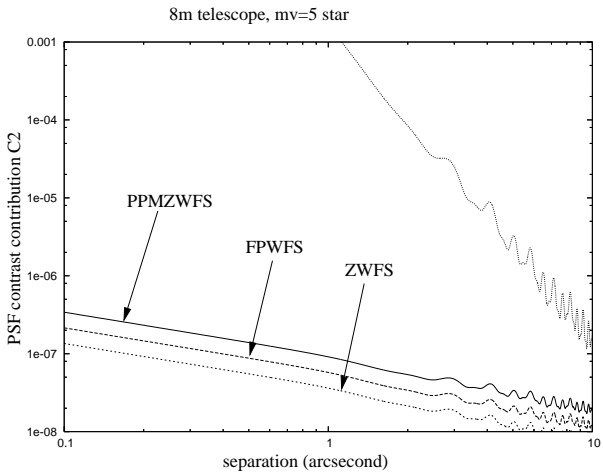}
\caption{\label{fig:C2phase8m} PSF contrast components $C_2$ obtained with PPMZWFS, FPWFS and ZWFS. The top curve shows the PSF contrast $C_0$ corresponding to uncorrected turbulence phase aberration. Parameters used for this simulation are listed in table \ref{tab:simul_param}.}
\end{figure}

Figure \ref{fig:C2phase8m} shows the PSF contrast component $C_2$ obtained with a FPWFS.

\subsection{Sensitivity of an ``Ideal'' WFS}
\label{sec:WFS_ideal}
Using the $(U,S_c)$ representation of WFSs given in appendix \ref{app:unif}, the unitary matrix $U$ can be optimized for the wavefront sensing of a pure sine wave phase error of fixed frequency $\vec{f}$ by finding the smallest possible value of $\beta_p$. Random WFSs can be built and the corresponding value of $\beta_p$ computed using the equations detailed in appendix \ref{app:wfsw}. The quantities $I_k$ measured are light intensities (number of photons), and the associated noise is $\sqrt{I_k}$. The numerical simulation results show that the value of $\beta_p$ obtained is then independant of the size of the unitary matrix used to represent WFSs (as long as this matrix is larger than 3x3) :
\begin{equation}
\label{equ:phn}
\beta_p = 1.
\end{equation}

Theoretically, an ``optimal'' WFS should therefore be able to have a sensitivity to photon noise $\beta_p = 1$. This result does not however insure that a WFS which can satisfy this requirement simultaneously for all spatial frequencies exists, as the above simulation was performed for a single spatial frequency. The results obtained in this work show that only the CWFS reaches this optimal sensitivity, but only for a single value of the spatial frequency $f$. Understanding how the CWFS achieves this result might allow the design of an ``Ideal'' WFS.

When detection is performed in the pupil plane, the goal of the WFS is to transform a phase aberration into a light modulation. I now consider two symmetric focal plane speckles of amplitude $a$ (relative to the central peak amplitude) and phases $\phi_1$ and $\phi_2$ (relative to the phase of the central peak). Fourier transform of this speckle pair yields pupil plane modulation amplitudes of 
\begin{equation}
\label{equ:Ma}
M_a = 2a \cos\left(\frac{\phi_1 + \phi_2}{2}\right)
\end{equation}
for amplitude and 
\begin{equation}
\label{equ:Mp}
M_p = 2a \sin\left(\frac{\phi_1 + \phi_2}{2}\right)
\end{equation}
for phase. As described by equation \ref{equ:syms}, a pupil plane sine-wave phase aberration of amplitude $\psi$ (in radian) and phase $\theta$ produces two focal plane symmetric speckles of amplitude $\psi/2$ and phases $\phi_1 = \pi/2-\theta$ and $\phi_2 = \pi/2+\theta$. Equations \ref{equ:Ma} and \ref{equ:Mp} confirm that, if the phase and amplitude are left unchanged in the focal plane, these 2 speckles correspond to a pure phase error in the pupil plane ($M_a = 0$). At its optimal spatial frequency, the CWFS adds $\pi/2$ to the phase of each speckle, resulting in $M_a = 2a$ and $M_p = 0$. In this particular case, the 2 speckles are interfering constructively together in the pupil plane. This level of amplitude modulation is impossible to reach if the 2 speckles are optically separated (Pyramid WFSs, FPWFS) and explains why only the CWFS can be ``optimal'' ($\beta_p=1$) for a spatial frequency. Its only drawback is that the pupil plane phase offset is $\pi/2$ only at a one value of the angular separation.

\begin{figure}[htb]
\hspace{-0.15in}\includegraphics[scale=0.32]{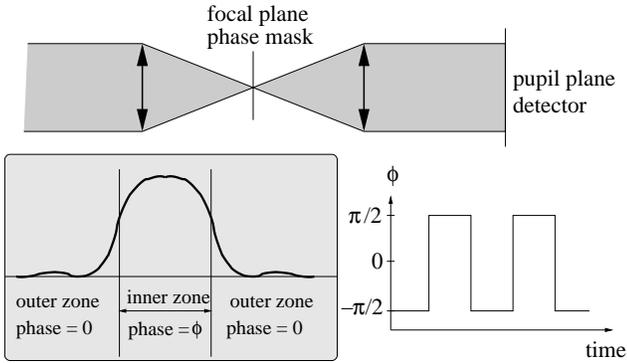}
\caption{\label{fig:idealWFS} Schematic representation of the Zernike WFS (ZWFS), the WFS with the minimum sensitivity to photon noise.}
\end{figure}

In order to build an ``optimal'' WFS, the focal plane offset would need to be $\pi/2$ at all separations. The most practical solution is to change the phase of the PSF core by $\pi/2$ and $-\pi/2$ alternatively, as shown in figure \ref{fig:idealWFS}. In closed-loop operation in an AO system, the phase offset does not need to be achromatic, but should be approximately $\pm \pi/2$ for minimum sensitivity (the phase offset in a CWFS is also not achromatic).
This WFS was originally developped for microscopy \citep{zern34}, and is named Zernike Phase Contrast Wavefront Sensor (ZWFS) in this paper. More recently, this WFS has been suggested for ground-based AO systems \citep{bloe03a}, because it offers a direct phase measurement (as opposed to wavefront slope for SHWFS). 
The phase mask should be small (size $\approx \lambda/d$) if the contrast needs to be optimized very close the PSF core.
In broadband, a speckle might be outside the phase mask in the red end of the band but inside it in the blue end. The sensitivity $\beta_p$ is then more than 1 in an intermediate transition region between the inner part of the PSF ($\beta$ is infinite) and the outer part of the PSF ($\beta_p = 1$). A Wynne corrector \citep{wynn79,rodd80}, which magnifies the pupil by a factor propotional to wavelength, may be used to avoid this effect. This device was originally developed to increase the spectral bandwidth of speckle interferometry, and has been successfully used on the sky \citep{bocc98}.

The ZWFS is highly sensitive ($\beta_p =1$, everywhere except possibly in the central core of the PSF if the mask is large) and quite achromatic, but has limited dynamical range: it is ideal when used after a low-order first stage AO system. 

Figure \ref{fig:C2phase8m} shows the PSF contrast component $C_2$ obtained with a ZWFS.

\subsection{Discussion}
\label{sec:wfs_beta_disc}
\begin{deluxetable}{lccccccc}
\tabletypesize{\tiny}
\tablecaption{\label{tab:WFS_comp} Comparative table of WFSs}
\tablewidth{0pt}
\startdata
 & SHWFS & CWFS & FPYRWFS & MPYRWFS & PPMZWFS & FPWFS & ZWFS\\
\hline
\hline
$\beta_p$ & $>2$ & $\geq 1$ & $\sqrt{2}$ & $\geq 2$ & $2$ & $\sqrt{2}$ & 1\\
\hline
$\beta_a$ & 1 & $\sqrt{\frac{1}{1-1/{\beta_p}^2}}$ & $\sqrt{2}$ & $\geq 1$ & 1 & $\sqrt{2}$ & $\infty$\\
\hline
Optimal sep. set by & \# lenslets & defocus & - & modul. rad. & - & - & -\\
\hline
Noise propagation & high & high & low & high & low & low & low\\
\hline
Achromaticity & good & good & good & good\tablenotemark{a} & good\tablenotemark{c} & good\tablenotemark{a} & good\\
\hline
Aliasing & high & high & moderate & moderate & moderate & none & moderate\\
\hline
Solutions to aliasing\tablenotemark{b} & SF & DS,SF,OF & DS,SF,OF & DS,SF,OF & DS,SF,OF & - & DS,SF,OF\\
\hline
Dynamical range & high & high & low & high & low & low & low \\
\hline
Elements & lenslets & pixels & pixels & pixels & pixels & pixels & pixels\\
\hline
Detectors per element & $\geq 4$ & $1$ & $4$ & $4$ & $2$ & $>1$\tablenotemark{d} & $1$\\
\enddata
\tablenotetext{a}{With the use of a Wynne corrector.}
\tablenotetext{b}{Aliasing can be reduced by higher detector sampling (DS), spatial filtering in the focal plane (SF), or an optical anti-aliasing filter (OF).}
\tablenotetext{c}{Requires achromatic phase shifts.}
\tablenotetext{d}{Set by focal plane pixel scale.}
\end{deluxetable}

Table \ref{tab:WFS_comp} summarizes the results obtained previously and lists $\beta_{p}$ and $\beta_a$ for the 7 WFSs compared in this study. 

\subsubsection{Sensitivity for OPD measurement}
\begin{figure}[htb]
\hspace{-0.0in}\includegraphics[scale=0.32]{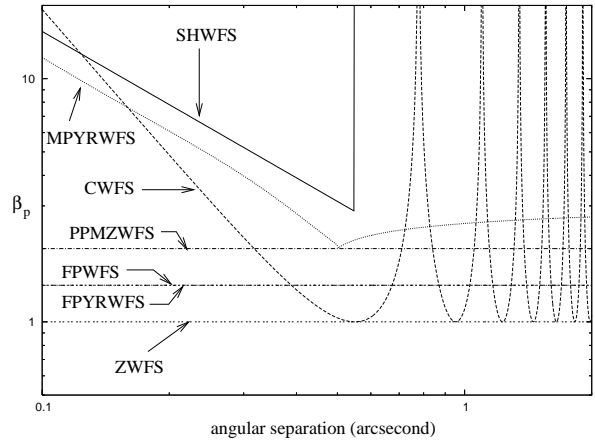}
\caption{\label{fig:betacomp} Value of $\beta_p$ as a function of angular separation for the WFSs compared in this study. The WFSs were optimized for a separation of 0\farcs5. For the SHWFS, $r_0 = 0.2m$ and $\lambda_0 = 0.5\mu m$.}
\end{figure}
\begin{figure}[htb]
\hspace{-0.0in}\includegraphics[scale=0.32]{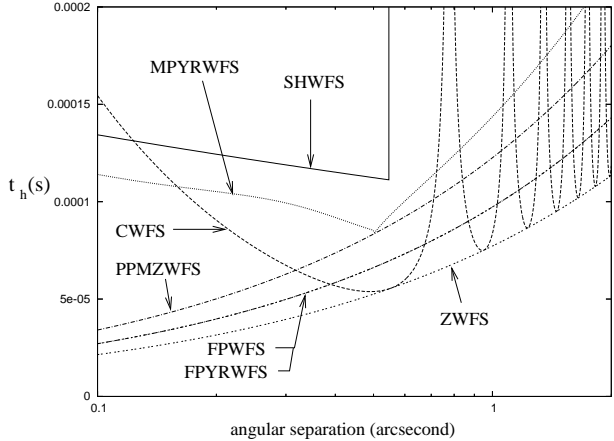}
\caption{\label{fig:t0comp} Value of $t_h$, the optimal sampling, as a function of angular separation for the WFSs compared in this study ($m_v=5$ source).}
\end{figure}
For some WFSs (SHWFS, CWFS and MPYRWFS) $\beta_p$ reaches its minimum at a given distance from the optical axis, and increases closer to the PSF core: these WFSs suffer from the noise propagation effect (low sensitivity to low-order modes due to photon noise). The other WFSs (FPYRWFS, PPMZWFS and FPWFS) maintain a constant value of $\beta_p$ at all separations: noise propagation is low, and low-order terms can be corrected efficiently at the same time as high-order terms. For most WFSs, correcting for amplitude+phase instead of only phase increases $\beta_p$, with the exception of the SHWFS, which measures amplitude ``for free'' (photometry within each subaperture). 
As will be shown in \S\ref{sec:contrastanalytical}, in the photon-noise limited regime, the PSF contrast achieveable by an AO system varies as ${\beta_p}^{4/3}$ (equation \ref{equ:C2}): the differences shown in figure \ref{fig:betacomp} are therefore important. For example, the contrast can be 1.6 times better in a CWFS-base or ZWFS-based AO system than in a FPWFS-based or FPYRWFS-based system. The ``ideal'' ZWFS outperforms the PPMZWFS by one magnitude in contrast. In the example shown in figure \ref{fig:betacomp}, the SHWFS, even perfectly tuned for optimal PSF contrast at 0\farcs5, produces a PSF contrast 4.3 worse than a ZWFS. As shown in figure \ref{fig:t0comp}, the sampling frequency required to reach optimal performance on a bright source is especially high at low spatial frequency for the most efficient WFSs.

Non-common path errors can limit the achievable contrast, and are almost unavoidable in pupil-plane WFSs (all WFSs except for FPWFS). Very accurate calibration is then required, and can be obtained by focal plane phase diversity. The FPWFS is immune to this effect if the wavefront sensing and scientific focal planes are shared, which is likely to be the case for visible coronagraphic imaging of extrasolar planets from space (TPF mission). However, on ground-based AO systems, wavefront sensing in the visible and scientific imaging in the near-IR is often prefered for scientific and technological (detectors) reasons.

All WFSs studied in this paper have good achromaticity and can be used in broadband light. The SHWFS, CWFS and FPYRWFS are naturally achromatic, while other WFSs require either achromatic phase shifters (PPMZWFS and ZWFS) or a Wynne corrector (FPWFS).

\subsubsection{Sensitivity for scintillation measurement}
The steps to compute $\beta_a$ are not detailed in this work, but comparison with the computation of $\beta_p$ reveals that $\beta_a = 1$ if all the light is used to image the pupil. From this result, $\beta_a$ can be easily estimated for all WFSs considered in this study.

In this work, I choose to adopt $\beta_a = 1$ for all WFSs in subsequent numerical simulations. While this is exact for the SHWFS, which does not optically modify the light intensity in the pupil plane, this is not true for most other WFSs. For example, in the CWFS, $1/{\beta_a}^2 + 1/{\beta_p}^2 = 1$ (equations \ref{equ:SigmaCVp}, \ref{equ:Ma} and \ref{equ:Mp}): at the optimal angular separation (defined by $\beta_p = 1$) the CWFS is insensitive to scintillation ($\beta_a = \infty$). If $\beta_a$ is high and $C_3 \gg C_2$, then a fraction of the total flux (or, equivalently, time) needs to be allocated to scintillation sensing, which is performed most efficiently by imaging of the pupil. For example, in the CWFS, a fraction of the time is spent at $dz = 0$ (no defocus in the focal plane). This sharing of the photons increases $C_2$ and decreases $C_3$ until $C_2+C_3$ is minimal.

However, as shown in \S\ref{sec:perf_disc}, $C_1<C_0$ within the central arcsecond: OPD aberrations are stronger than scintillation at low spatial frequencies. Since both terms are moving in front of the telescope with the same speed $v$, the post-correction scintillation residual $C_3$ can be made comparable to post-correction OPD residual $C_2$ by allocating a small fraction of the incoming photons to scintillation measurement. The PSF contrasts obtained with the approximation $\beta_a=1$ are therefore only slightly optimistic within the central arcsecond: $\beta_p$ sets the value of $C_2+C_3$, not $\beta_a$.

Beyond $\alpha = $ 1\arcsec, however, $C_0 \approx C_1$, and if a WFS is characterized by $\beta_a = \infty $, $\beta_p = 1$, half of the photons should be allocated to pure scintillation measurement. This would result in $\beta_a = \beta_{ap} = \sqrt{2}$, which would produce contrasts $C_2$ and $C_3$ equal to $2^{2/3} \approx 1.6$ the values obtained with the optimistic approximation $\beta_a=1$. The maximum error made by the $\beta_a = 1$ approximation is therefore a factor 1.6 on $C_2$ and $C_3$, and can only occur at large angular separation ($\alpha>$1\arcsec) with the CWFS and the ZWFS.

\section{Contrast performance}
\label{sec:perf_disc}

\subsection{Parameters adopted for numerical simulations}
Table \ref{tab:simul_param} lists the default parameters adopted in this work for numerical simulations. The atmospheric parameters correspond to conditions frequently encountered atop Mauna Kea, Hawaii. The weigths and altitudes of the turbulence layers are derived from 4 nights of MASS and Scidar measurement atop Mauna Kea \citep{toko05}. The photometric zero point of the WFS (corresponding to an equivalent bandpass of $0.1 \mu m$) is representative of existing WFSs. 

Through the paper, some of these parameters are modified to evaluate the contrast performance of a system which departs from this default configuration: wavefront sensing and imaging wavelength in \S\ref{sec:choice_lambda} and atmospheric parameters in \S\ref{sec:obs_site}. The contrast performance can also easily be derived for telescopes larger than 8m: since the contrast limits $C_0$ to $C_6$ are all proportionnal to $1/D^2$, the overall contrast for all WFSs is proportionnal to $1/D^2$.

\subsection{Relative contribution of contrast limits components in conventional AO.}
\begin{deluxetable}{lll}
\tabletypesize{\small}
\tablecaption{\label{tab:simul_param} Default simulation parameters}
\tablewidth{0pt}
\startdata
\hline
Telescope diameter & $D$ & 8 m \\
Detection wavelength & $\lambda_i$ & 1.6 $\mu$m\\
Source brightness & $m_v$ & 5\\
\hline
WFS wavelength & $\lambda$ & 0.55 $\mu$m\\
WFS bandpass & & 0.1 $\mu$m\\
WFS quantum efficiency & & 1.0\\
WFS zero point (counts for $m_v=0$ source) & & $9.74\:10^{9}$ cnt/s/m$^2$\\
WFS sensitivity to OPD & $\beta_p$ & 1.0\\
WFS sensitivity to scintillation & $\beta_a$ & 1.0\\
\hline
Fried parameter & $r_0$ at 0.5 $\mu$m & 0.2 m (0.5'' seeing)\\ 
Wind speed & $v$ & 10 m/s\\
Number of layers &  & 6\\
Layer 1 altitude / $C_n^2$ fraction &  & 0.5km / 0.2283\\
Layer 2 altitude / $C_n^2$ fraction &  & 1.0km / 0.0883\\
Layer 3 altitude / $C_n^2$ fraction &  & 2.0km / 0.0666\\
Layer 4 altitude / $C_n^2$ fraction &  & 4.0km / 0.1458\\
Layer 5 altitude / $C_n^2$ fraction &  & 8.0km / 0.3350\\
Layer 6 altitude / $C_n^2$ fraction &  & 16.0km / 0.1350\\
effective altitude $(\mu_2/\mu_0)^{1/2}$ & $z_2$ & 7.7km\\
effective altitude $(\mu_{5/3}/\mu_0)^{3/5}$ & $\bar{h}$ & 7.1km\\
\enddata
\end{deluxetable}

\begin{figure}[htb]
\includegraphics[scale=0.32]{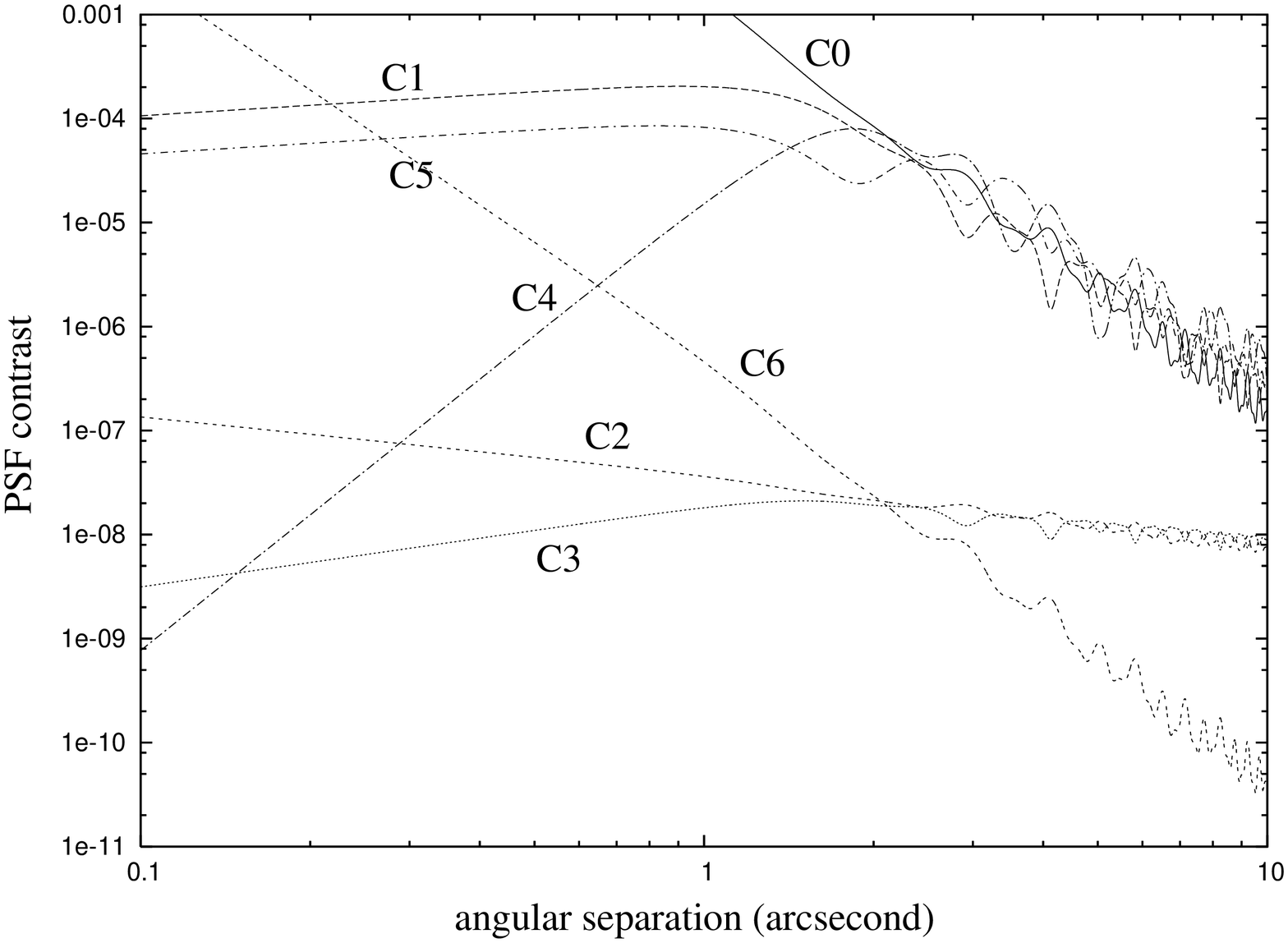}
\caption{\label{fig:C01234} Contrast limits imposed by the uncorrected atmospheric turbulence (C0 and C1), corrected atmospheric turbulence (C2 and C3), chromatic effects (C4, C5, and C6) for a 8m telescope and a $m_v=5$ source. See text for details.}
\end{figure}

Figure \ref{fig:C01234} shows the relative contributions of $C_0$, $C_1$, $C_2$, $C_3$, $C_4$, $C_5$ and $C_6$ when an ideal WFS (ZWFS) is used on a bright ($m_V=5$) star. According to the results obtained in \S\ref{sec:WFSs}, this wavefront sensing scheme is the most sensitive, and other WFSs will show higher values of $C_2$. The main parameters of the simulation are listed in table \ref{tab:simul_param}, and are used through this work unless otherwise specified. 
Chromatic effects introduced by either the refraction index of air (component $C_6$) or Fresnel propagation through the atmosphere ($C_4$ and $C_5$) can have a strong impact on the PSF contrast. One very important result from this study is that $C_0$, $C_1$, $C_4$ and $C_5$ are all comparable beyond 2\arcsec. Similarly, $C_2$ and $C_3$ are comparable beyond about 2\arcsec. The implications of this result are now discussed separately for AO systems correcting only OPD and AO systems correcting OPD and scintillation. For now, I choose to limit this discussion to AO systems performing wavefront sensing in the visible and imaging in the near-IR, as choices of wavelengths will be discussed in \S\ref{sec:choice_lambda}.

\begin{itemize}
\item{{\bf OPD correction with AO : } In an OPD-only AO correction ($C=C_1+C_2+C_4+C_6$), the uncorrected scintillation $C_1$ dominates by far the achievable PSF contrast within the central 2\arcsec, and limits it to approximately $10^{-4}$ to $2\:10^{-4}$ within the central arcsecond. The term due to photon noise and time lag, $C_2$ is much lower, at about $10^{-7}$ for this bright ($m_v=5$) source. The effect of photon noise would become dominant only for $m_V>13$ with a high-sensitivity WFS ($\beta_p=1$). The OPD chromaticity term $C_4$ due to Fresnel propagation is small close to the PSF center, but is rapidly increasing with angular separation, and is comparable to scintillation $C_1$ at 2\arcsec and beyond.}
\item{{\bf OPD+scintillation correction with AO : } With an AO system correcting both OPD and scintillation ($C=C_2+C_3+C_4+C_5+C_6$), the term $C_5$ due to the chromaticity of scintillation limits the PSF contrast to slightly better than $10^{-4}$ in the central 2\arcsec. Scintillation, if measured at $\lambda \approx 0.5\mu m$, cannot be well corrected for at $\lambda_i \approx 1.6 \mu m$. The improvement to the PSF contrast brought by correction of scintillation with the AO system is quite modest (about a factor of 2).}
\end{itemize}
Beyond about 2\arcsec, $C \approx C_0$, and ``classical'' AO (wavefront sensing in visible, imaging in near-IR), even with scintillation correction, cannot improve the PSF contrast: there is no use to increase the number of elements beyond this limit.

\subsection{Choice of wavefront sensing and imaging wavelengths}
\label{sec:choice_lambda}

The number of photons available for wavefront sensing is a function of spectral type, and should be maximized to reduce the PSF contrast $C_2$.
In this section, I consider the number of photons available for wavefront sensing to be independant of $\lambda$, which is a good approximation for a spectral type G2 and a fixed spectral bandwidth ($\delta\lambda/\lambda$ constant).

\begin{figure*}[htb]
\includegraphics[scale=0.3]{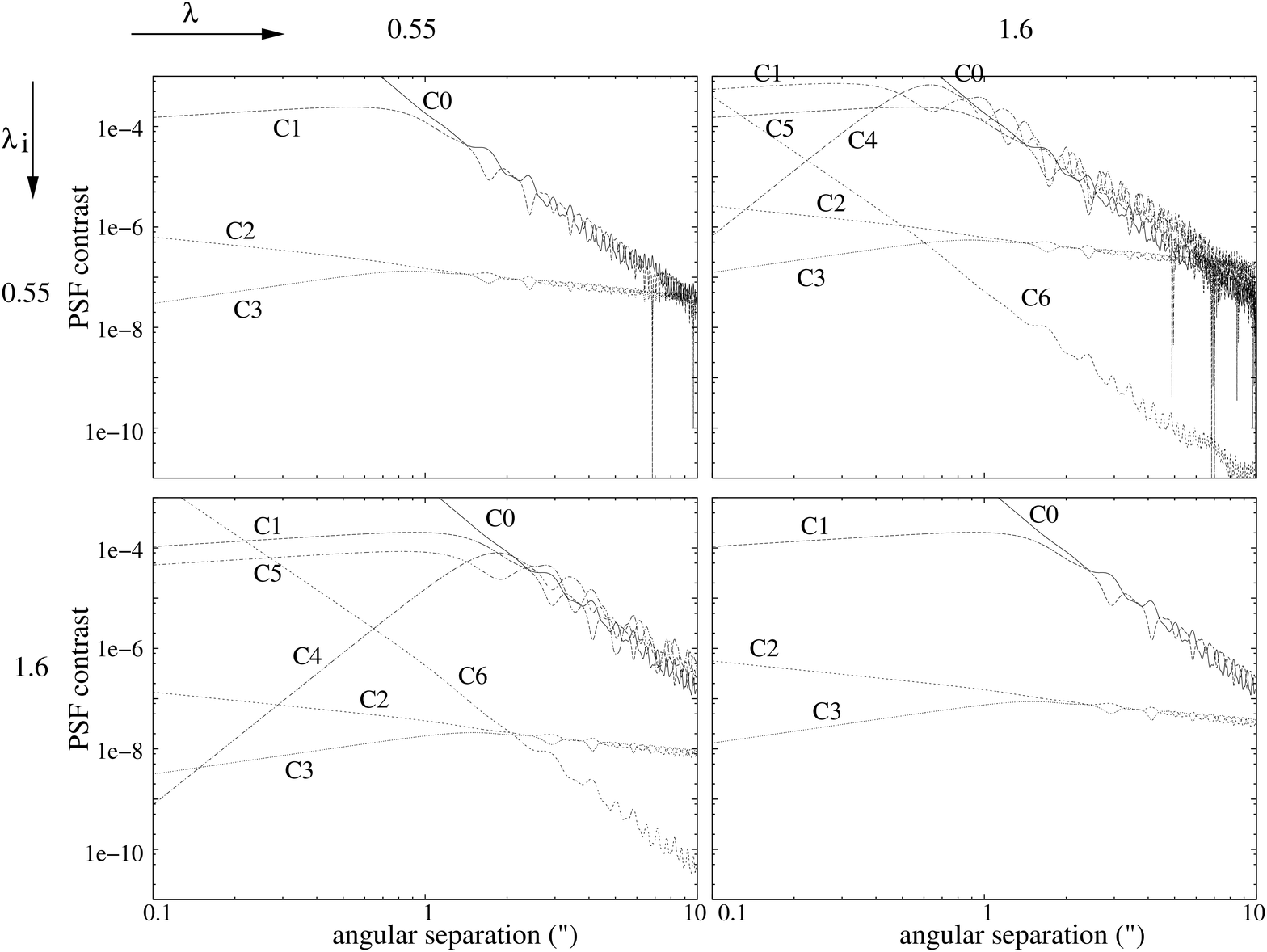}
\caption{\label{fig:lambdachoice} PSF contrast components $C_0$ to $C_6$ as a function of angular separation for different combinations of WFS wavelength $\lambda$ (shown at the top, in micron) and imaging wavelength $\lambda_i$ (shown on the left, in micron).}
\end{figure*}

Figure \ref{fig:lambdachoice} illustrates how $\lambda$ and $\lambda_i$ affect PSF contrast components $C_0$ to $C_6$. 
When $\lambda = \lambda_i$, chromatic terms ($C_4$, $C_5$ and $C_6$) disappear, and the PSF contrast is driven by WFS photon noise through $C_2$ and $C_3$ (for an AO system correcting OPD and scintillation) or $C_1$ (for an AO system correcting only OPD). In all configurations, $C_6$ has a negligible impact on PSF contrast beyond 0\farcs3.

\begin{figure}[htb]
\includegraphics[scale=0.32]{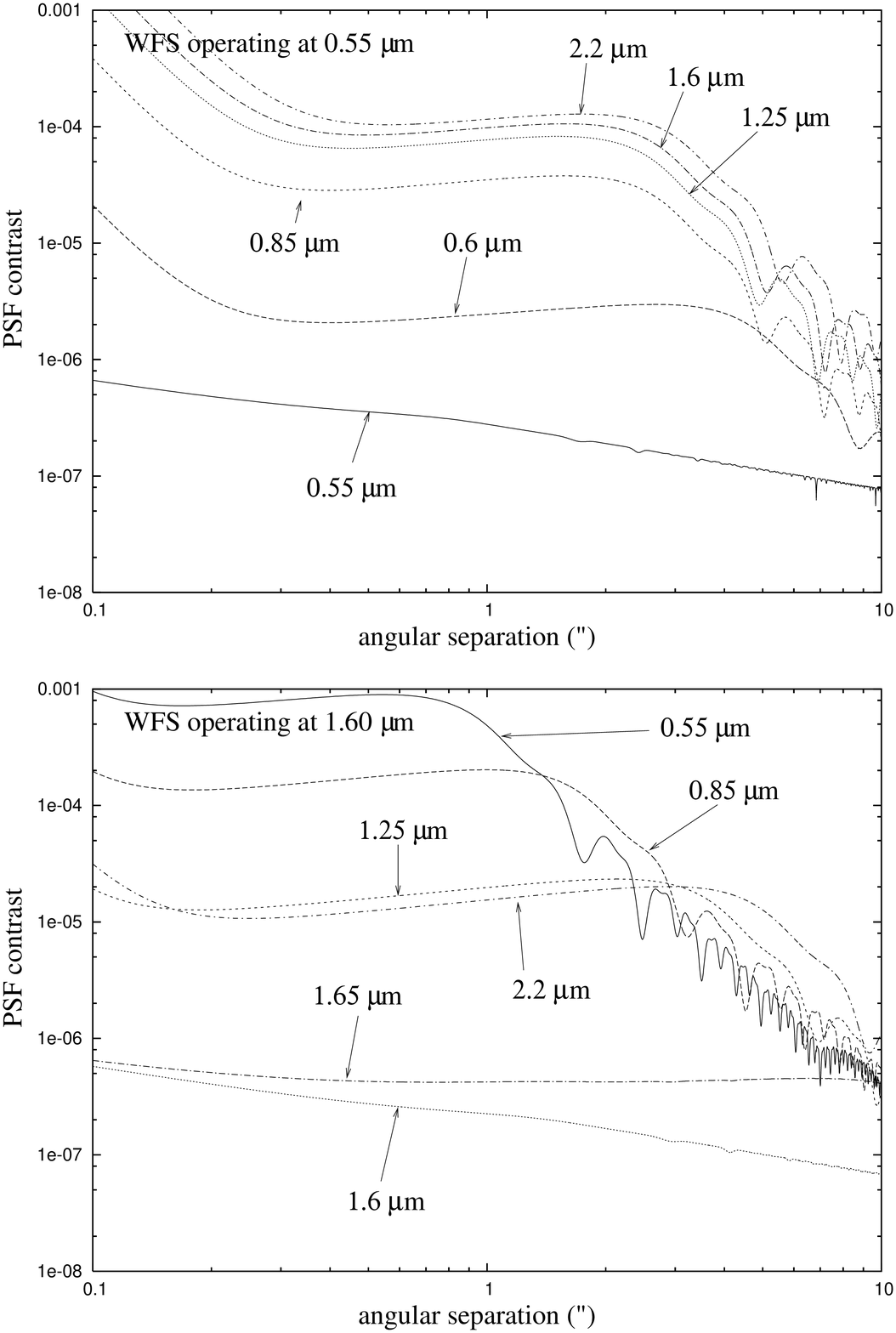}
\caption{\label{fig:AOCperf} PSF contrast for a WFS operating at $0.55 \mu m$ (top) and $1.6 \mu m$ (bottom) after AO correction of OPD and scintillation. The PSF contrast is shown at a function of angular separation at imaging wavelengths ranging from $0.55 \mu m$ to $2.2 \mu m$ in each case.}
\end{figure}

These results are combined in figure \ref{fig:AOCperf}, which shows the achievable PSF contrast as a function of imaging wavelength when the WFS wavelength is fixed. An AO system correcting both OPD and scintillation is considered in this figure, with an ``ideal'' WFS operating in the visible (top) or in the near-IR (bottom). When $\lambda=\lambda_i$, the photon-noise driven PSF contrast is between $10^{-6}$ and $10^{-7}$ in the central arcsecond in both cases. The PSF contrast degrades very rapidly as $\lambda_i$ becomes different from $\lambda$. In conventional AO, where $\lambda \approx 0.55 \mu m$ and imaging is performed in the near-IR, the PSF contrast, dominated by chromatic effects, is limited to a few times $10^{-5}$ in the central arcsecond. In order to approach the photon-noise limit ($C_2+C_3$), imaging needs to be performed within about $0.05 \mu m$ of the wavefront sensing wavelength. When observing a bright source, figure \ref{fig:AOCperf} illustrates that the wavefront sensing wavlength should be chosen equal to the imaging wavelength.

This statement is at variance with the common practice of combining visible wavefront sensing and near-IR imaging to yield the best possible contrast. Figure \ref{fig:lambdachoice} does indeed show that this is the optimal choice if chromatic effects are ignored. In current AO systems, the WFS detector and sensitivity $\beta_p$ are sub-optimal (the perfect example is a SHWFS with finite readout noise CCD), and errors are dominated by photon noise except for the brightest sources: in this regime, figure \ref{fig:lambdachoice} shows that visible $\lambda$ and near-IR $\lambda_i$ is optimal. Moreover, finite number of actuators, lack of aliasing mitigation scheme and non-common path errors set a limit to the PSF contrast even for bright sources: the effect of these OPD errors on the PSF contrast is mitigated by increasing $\lambda_i$. For these reasons, most current AO systems do not achieve a $10^{-4}$ PSF contrast in the central arcsecond, and are therefore not dominated by chromatic effects. However, an AO system designed to maximize PSF contrast (using an efficient WFS and an aliasing mitigation scheme) would be dominated by chromatic effects if $\lambda \neq \lambda_i$.

If a visible WFS is used for imaging in the near-IR, the chromatic components $C_4$ and $C_5$ are fixed phase and amplitude screens moving in front of the telescope's pupil with speed $v$. The coherence time of the corresponding speckles is therefore long, unlike the fast residual speckles due to time lag and photon noise (contributions $C_0$ and $C_1$). These slow speckles are very detrimental to the final detection limit, as they require long exposure times to average into a smooth continuous background.

\subsection{Observing site}
\label{sec:obs_site}

\begin{deluxetable}{lcccccc}
\tabletypesize{\small}
\tablecaption{\label{tab:obs_site} Observing sites characteristics}
\tablewidth{0pt}
\startdata
Site & Seeing & $r_0$ & $\tau_0$ & $v$ & $\theta_0$ & $z_2$ \\
     & (\arcsec) & (m) & (ms) & (m/s) & (\arcsec) & (km)  \\
\hline
\hline
This work & 0.5 & 0.2 & 6.3 & 10 & 1.5 & 7.7\\
\hline
A & 0.58 & 0.17 & 2.7 & 19.8 & 1.9 & 6.3 \\
B & 0.35 & 0.29 & 5.4 & 16.9 & 3.8 & 5.4\\
\hline
C & 0.27 & 0.37 & 7.9 & 14.7 & 5.7 & 4.6 \\
D& 0.10 & 1.0 & 20 & 15.7 & 8.5 & 8.3 \\
\hline
E & 0.74 & 0.14 & 3.3 & 13.3 & 2.6 & 3.8\\
F & 0.50 & 0.2 & 6.9 & 9.1 & 3.6 & 3.9\\
\hline
\enddata
\end{deluxetable}

\begin{figure}[htb]
\includegraphics[scale=0.32]{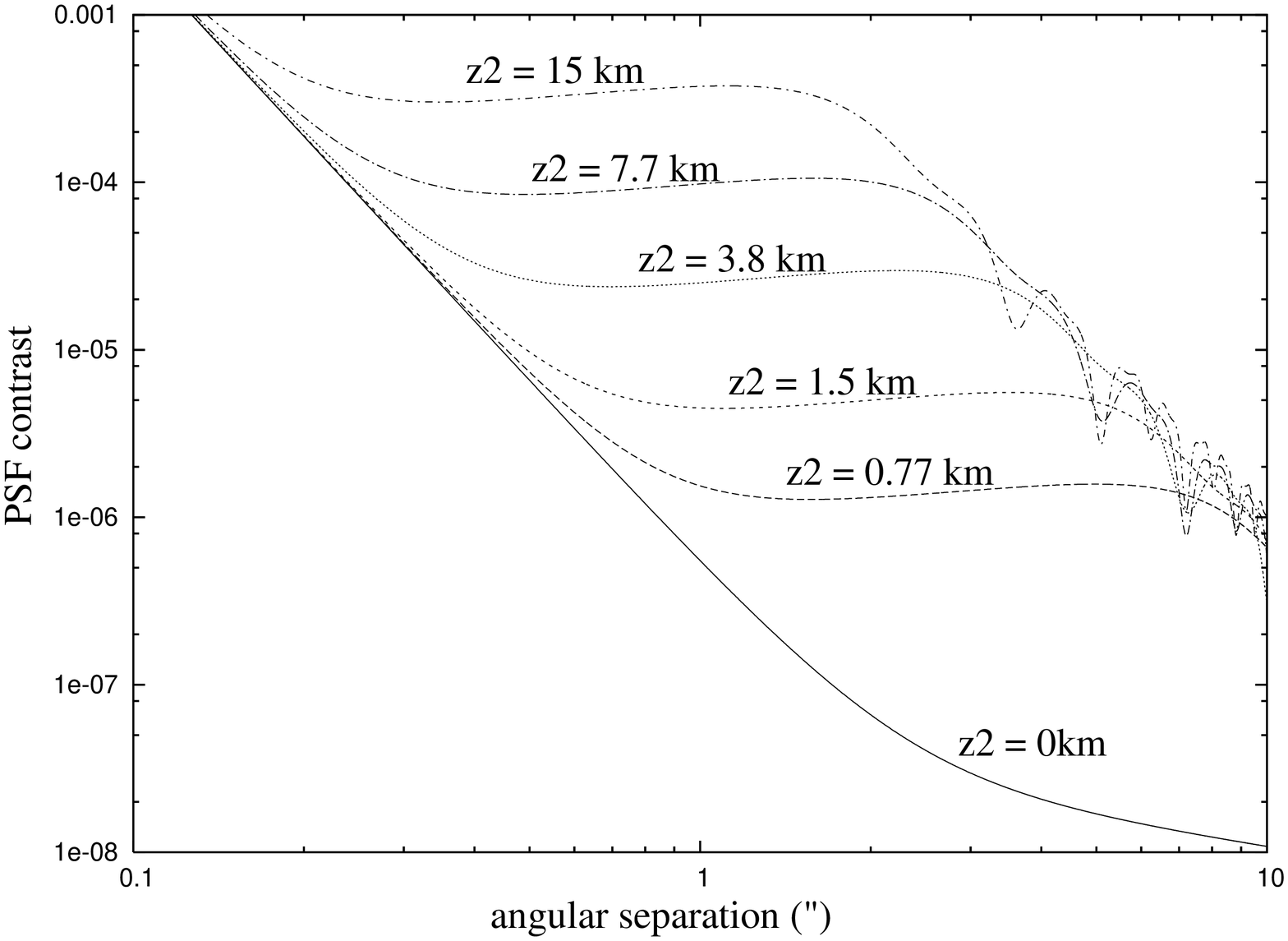}
\caption{\label{fig:site_z} Effect of the effective turbulence altitude $z_2$ on the PSF contrast. Parameters other than $z_2$ for this simulations are given in table \ref{tab:simul_param}.}
\end{figure}

If $\lambda \neq \lambda_i$ (contrast dominated by chromaticity effect), the PSF contrast $C$ is dominated by $C_5$ as illustrated in figures \ref{fig:lambdachoice} and \ref{fig:AOCperf}.  Equation \ref{equ:C5} for small angular separations (within the central arcsecond) leads to:
\begin{equation}
C \propto {z_2}^2 \: {r_0}^{-5/3}
\end{equation}
where 
\begin{equation}
z_2 = \sqrt{\frac{\int C_n^2(z) \: z^2 \: dz}{\int C_n^2(z) \: dz}}.
\end{equation}
The contrast is then independant of wind speed $v$. In this regime, multiplying $z_2$ by 0.6 (40\% lower altitude turbulence) is equivalent to improving the seeing by a factor 2. The high importance of turbulence heigth on PSF contrast is illustrated in figure \ref{fig:site_z}, where the turbulence profile given in table \ref{tab:simul_param} has been scaled in altitude to modify $z_2$. All other parameters of for this simulation are given by table \ref{tab:simul_param}.

If $\lambda \approx \lambda_i$ (chromaticity effects are small), the PSF contrast is dominated by $C_2$, as shown in figure \ref{fig:lambdachoice}. At small angular separations, 
\begin{equation}
C \approx {r_0}^{-5/9} \: v^{2/3}.
\end{equation}
The contrast is then independant of the atmospheric turbulence profile.

\begin{figure}[htb]
\includegraphics[scale=0.32]{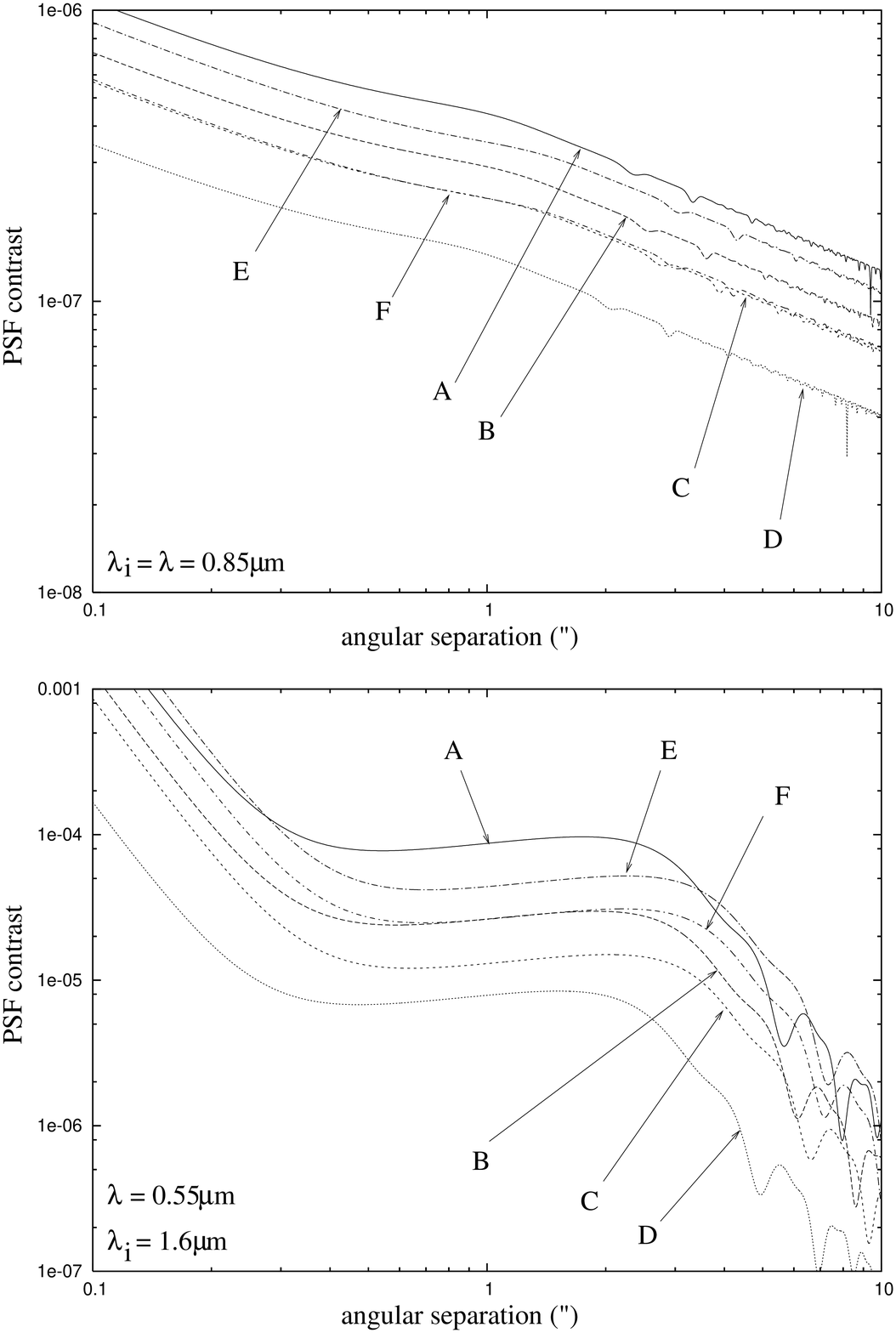}
\caption{\label{fig:perfsite} PSF contrast for different observing sites. On the top panel, $\lambda=\lambda_i=0.85 \mu m$. On the bottom panel, $\lambda=0.55 \mu m$ and $\lambda_i = 1.6 \mu m$. Parameters other than $r_0$, $v$ and $z_2$ for this simulations are given in table \ref{tab:simul_param}.}
\end{figure}

Figure \ref{fig:perfsite} shows the PSF contrast for the observing conditions listed in table \ref{tab:obs_site} for $\lambda=\lambda_i=0.85 \mu m$, and for $\lambda=0.55 \mu m$, $\lambda_i = 1.6 \mu m$. Parameters not listed in table \ref{tab:obs_site}, such as telescope diameter, were taken from table \ref{tab:simul_param}. Atmospheric conditions A and B in table \ref{tab:obs_site} were derived from measurements atop Mauna Kea \citep{naoj04,raci95}, where A is representative of median conditions, while B corresponds to excellent conditions (roughly 10th percentile). Models C and D corresponds to respectively median and 10th percentile conditions measured by \citet{lawr04} at Dome-C, Antartica. Since these measurements were acquired over a short timescale (22 nights within a 50 days period), and were insensitive to low altitude (less than 30m) turbulence, they might not accurately represent the site's atmospheric conditions. Models E and F are representative of median and 10th percentile conditions on Cerro Parannal \citep{eso04}.

The atmospheric turbulence outer scale $L_0$, which was not included in the model used in this study, is usually larger than 10m, and has therefore no direct effect on the PSF contrast beyond 0\farcs05 in the near-IR.

\section{The case for focal plane wavefront sensing}
\label{sec:fpwfs_layout}

The results obtained in \S\ref{sec:perf_disc} suggest that wavefront sensing and imaging should be performed at the same wavelength. The PSF contrast is then driven by the sensitivity of the WFS, and can reach up to $10^{-7}$ in visible/near-IR on a 8m telescope in the central arcsecond for a bright source. To achieve this level of contrast, a high sensitivity WFS should be chosen, and non-common path errors need to be carefully calibrated.

The FPWFS offers high sensitivity (only surpassed by the ZWFS) and does not suffer from aliasing or non common path errors (if the same focal plane is used for scientific imaging and wavefront sensing). If amplitude and OPD errors are comparable and both need to be measured, FPWFS is in fact as sensitive as the ``ideal'' ZWFS. Focal plane wavefront sensing is therefore an extremely attractive solution for high contrast ground-based AO, and I show in this section how it could be implemented.

In order to estimate the complex amplitude in the focal plane, a set of N reference waves are combined in the focal plane with the speckles. Following the notation adopted in \S\ref{sec:WFS_FP}, the complex amplitude of the reference wave $k$ is noted $(x_k,y_k)$ (real and imaginary parts) and is a function of the position in the focal plane.
As pointed out in \S\ref{sec:WFS_FP}, the amplitudes of the reference waves are not important, provided that they are larger than the amplitude of the speckles that need to be sensed, and a minimum of 2 waves are needed to reach optimal performance. If only 2 waves are used, their phases should be offset by $\pi/2$ ideally. As more reference waves are used, the number of nearly-optimal solutions increases and more flexibility in the choice of the reference waves is available.

The reference waves can be produced by phase-shifting of light extracted from the central part of the PSF in the focal plane \citep{ange03,codo04}. The interference between the speckle cloud and the reference waves is obtained through a Mach-Zender type interferometer with a beam splitter. This approach allows to create the optimal reference wave for focal plane wavefront sensing but requires additionnal optics.

A slightly less optimal, but optically simpler solution is to use the DM to produce the required reference waves. Each reference wave is created by a command sent on the DM: for example, moving a single actuator produces a reference wave of quasi-constant amplitude within the control region, and of phase given by the position of the actuator within the pupil. Provided that the influence functions of the DM and the behaviour of the coronagraph are known, the complex amplitude of the reference wave in the focal plane can be computed to a good accuracy. In a fast closed-loop system, the accuracy with which these reference waves are known does not need to be very high (about 10\% accuracy is sufficient) to be photon-noise limited.

Although the full optimization of the set of reference waves is beyond the scope of this work, I now show that reference waves well suited to focal plane wavefront sensing can be obtained with the DM :
\begin{itemize}
\item{{\bf Extent of reference wave in the focal plane.} The size of a reference wave in the focal plane is set by the size of an actuator of the DM. It is therefore possible to produce reference waves with relatively high amplitude across all of the diffraction control domain (DCD, the region in the focal plane for which the DM sampling is sufficient to suppress diffracted light).}
\item{{\bf Ability to set the amplitude of the reference wave.} Small amplitude waves can easily be created by moving a single actuator. Increasing the displacement of this actuator will increase the reference wave amplitude up to $1/N_{act}$ of the peak amplitude of the PSF for an unapodized pupil, where $N_{act}$ is the total number of actuators in the DM. For a PSF contrast better than $10^{-7}$ and a DM with more than $10^7$ actuators, actuators need to be moved in groups to produce reference waves of sufficient amplitudes.}
\item{{\bf Ability to produce an achromatic phase in the reference wave.} A small positive displacement of the center actuator of the DM produces a reference wave of achromatic phase $\pi/2$ in the focal plane ($-\pi/2$ for a negative displacement). The amplitude of this reference wave is however chromatic, but this effect will not seriously affect the system performance in a closed-loop system, as it is equivalent to multiplying the signal (speckle intensity modulation) by a wavelength-dependent gain of constant sign. The phase of reference waves produced by moving actuators other that the central actuator are also achromatic, provided that a Wynne corrector is used.}
\end{itemize}
Actuators near the edge of the pupil should preferably not be used to produce the focal plane reference waves, as the resulting phase would vary rapidly across the focal plane image.

\section{Coronagraphy}
\label{sec:coronagraphy}
The contrast limits derived in \S\ref{sec:WFSs} and \S\ref{sec:perf_disc} assumed that the only sources of scattered light are the wavefront errors (phase and amplitude) in the corrected beam. These results only apply to optical systems in which the static diffraction (Airy pattern on a circular aperture) is below the contrast levels derived in this work. In this section, I discuss the validity of this approximation, and briefly summarize the options available (coronagraphy) to insure that the limits derived in \S\ref{sec:WFSs} and \S\ref{sec:perf_disc} can be reached.

\subsection{Need for suppression of Airy pattern}
\begin{figure}[htb]
\hspace{-0.15in}\includegraphics[scale=0.32]{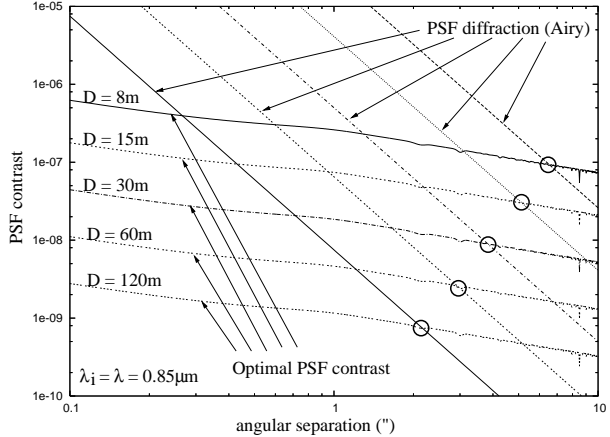}
\caption{\label{fig:contrast_Airy} PSF contrast with an optimal WFS ($\beta=1$) and Airy pattern diffraction for 8m, 15m, 30m, 60m and 120m diameter telescopes. The black circles mark the points where the theoretical PSF contrast achievable with a perfect coronagraph equals the Airy pattern diffraction contrast level.}
\end{figure}

Figure \ref{fig:contrast_Airy} shows that with an efficient WFS, diffraction associated with the Airy pattern is much stronger than light diffracted by residual wavefront errors at small angular separations. At some distance from the optical axis, both contributions are equal. Beyond this separation, coronagraphy is not required to reach the contrast level achievable by the AO system. As can be seen in figure \ref{fig:contrast_Airy}, this critical angular separation decreases as the telescope diameter increases. For reasonnable size telescopes (100m diameter or less), this critical separation is larger than 2\arcsec, and suppression of the Airy pattern is therefore required.
While light diffracted by residual wavefront errors is time-variable, and might not average nicely with time, the Airy diffraction pattern on the other hand is very stable and can be calibrated accurately. One might therefore wonder if relatively high levels of static diffraction features in the PSF are really detrimental to high contrast imaging. Static diffraction actually amplifies the time-variable speckles, an effect referred to as ``speckle pinning'' \citep{bloe01,bloe03,aime04}. This is due to the fact that the complex amplitudes of static and dynamic speckles add in the focal plane, and the light intensity $I$ measured is:
\begin{equation}
I = (A_s+A_d)^2 = I_s + 2\sqrt{I_s I_d} + I_d
\end{equation}
where $I_s$ and $I_d$ are the static and dynamic focal plane intensities, and the phase term between the 2 contribution has been ommited for simplicity. Assuming that $I_s$ is well known and can be perfectly subtracted, if $I_s>I_d$ (Airy patter is bighter than $C$), the ``speckle noise'' becomes dominated by $2\sqrt{I_s I_d}$ and is therefore $2\sqrt{I_s/I_d}$ stronger than it would be if $I_s=0$. In the example considered in figure \ref{fig:contrast_Airy}, at 0\farcs5 separation, adding a coronagraph reduces the ``speckle noise'' by factors 52, 26 and 19 on 8m, 30m and 60m telescopes respectively.

With focal plane wavefront sensing, the static diffraction of the telescope pupil can be treated just as atmospheric speckles, and can therefore be perfectly cancelled in half of the focal plane with a single DM. The DM phase is then used to cancel the static diffraction (Airy rings for example) created by the pupil intensity distribution, removing the need for a coronagraph, or at least lowering the requirement on its intrinsic performance. Since the DM phase is chromatic, but the pupil intensity is not, the resulting dark region of the PSF is chromatic \citep{codo04} and does not allow the use of a wide spectral band. \citet{yang04} however found that solutions with very low chromaticity exist with a square telescope pupil, and similar solutions might exist with circular telescope pupils. If successfull, this technique could be used instead of a coronagraph in narrow band imaging, and would be optically very simple, provided that the DM can produce the required phase functions: the solutions found by Yang and Kostinski require a large phase slope at the edge of the pupil. 

\subsection{Coronagraph/WFS combination}

At large angular separation the coronagraph does not need to attenuate the Airy pattern by a large factor (at most a factor 100 attenuation is required at 0\farcs5 according to figure \ref{fig:contrast_Airy}), and many suitable coronagraphic options are therefore available. 

When observation at small angular separation (a few $\lambda/d$) is required, the choice of the coronagraph is more critical. Coronagraphs with small IWA exist, but are often prone to:
\begin{itemize}
\item{(1) Sensitivity to tip-tilt and low order modes: This effect is problematic on coronagraphs for which small (less than $\lambda/d$) tip-tilt errors can scatter light at large outside the IWA \citep{rodd97,roua00,baud00}. The coronagraphic leaks are especially large if they are combined with a WFS having poor sensitivity to low-order modes, such as a CWFS or a high-order SHWFS. A FPWFS is preferable if the contrast at very small IWA needs to be optimized. Coronagraphs relying on pupil apodization manage to keep this effect small while offering small IWA \citep{kasd03,guyo05}.}
\item{(2) Chromaticity: Coronagraphs with small focal plane masks are sensitive to the wavelength-dependent PSF scale. Several designs have been proposed to mitigate \citep{soum03a,roua00} or solve \citep{baud00} this problem.}
\item{(3) Reduced throughput due to pupil and/or focal plane masks \citep{kasd03,kuch03,soum03}, or splitting of light \citep{baud00}: Lossless apodization can be used to avoid this problem \citep{guyo05}.}
\item{(4) Lower image quality: broader PSF due to apodization \citep{kasd03} or double images \citep{baud00}.}
\end{itemize}
These effects can be especially problematic if the WFS is placed after the coronagraph, as would likely be the case in a FPWFS-based system: chromaticity, reduced throughput and lower image quality would then effectively reduce the WFS signal-to-noise ratio and compromise the achievable contrast ratio.

\section{Conclusion}
A thourough comparison of the fundamental contrast limits of AO has shown that visible wavefront sensing doesn't allow high accuracy correction of near-IR wavefront aberrations: chromatic effects then limit the PSF contrast to $10^4$ to $10^5$ within the central arcsecond. Wavefront sensing should therefore be performed at the same wavelength as imaging to reach the contrast limit imposed by photon noise (about $10^{-6}$ to $10^{-7}$ in the central arcsecond).

An AO system optimized for high contrast, with wavefront sensing and imaging at the same wavelength, can still greatly benefit from a first stage AO correction with a shorter wavelength (visible) WFS:
\begin{itemize}
\item{If residual aberrations are small, a FPWFS can be used efficiently. This WFS offers unique advantages: no non-common path errors, no aliasing, high sensitivity. It therefore appears to be the ideal solution for high contrast AO.}
\item{If wavefront correction were perfect at the shorter wavelength, the chromatic residuals that the second WFS needs to measure would be small and relatively slow (same speed as the uncorrected turbulence). The PSF contrast achievable in this case could be better than the limits derived in this work, as more photons are used.}
\end{itemize}
This solution is especially attractive since low-noise fast visible detectors exist, while near-IR detectors currently offer lower performance. Theoretically, combining a fast high sensitivity visible WFS (preferably a ZWFS) with a slower near-IR WFS (preferably a FPWFS) is not as advantageous for red sources as it is for bluer (spectral type G or bluer) sources.


The PSF contrast estimates derived in this paper represent a limit which is hard to reach, as many optimistic hypothesis have been made: observation at zenith, perfect telescope and detectors, perfect DM, high system throughput, favorable atmospheric conditions, perfect coronagraph, bright $m_v=5$ source, no time delay for AO control. On a 8m telescope, PSF contrast up to $10^{-6}$ may be reached in the central arcsecond if deviation from these optimistic assumptions is minimal. Even on a 100m telescope, the corresponding contrast (about $10^{-8}$) is two orders of magnitude short of what is required to detect an Earth-size planet orbitting a solar-type star. Direct imaging of extrasolar planets is therefore bound to rely heavily on efficient calibration of the speckle noise (through differential imaging techniques for example).

\acknowledgements
The author is thankful to the referee, Ren\'e Racine, for his detailed and thorough look at this work, which led to many useful suggestions.

\appendix
\section{Computing the sensitivity of a WFS to photon noise for the measurement of a pure sine wave phase error}
\label{app:wfsw}
In this appendix, I define $\beta_p$, the quantitative measure of sensitivity to photon noise of a WFS, and show how it can be derived. The steps described here were used for each of the WFSs considered in this study to compute $\beta_p$ for a range of spatial frequencies.

The wavefront error given in equation \ref{equ:phiuv} can be represented in a 2D plane coordinate system by 
\begin{equation}
x_0 = \frac{2 \pi h}{\lambda} \cos(\theta)
\end{equation}
\begin{equation}
y_0 = \frac{2 \pi h}{\lambda} \sin(\theta)
\end{equation}
where $x_0$ and $y_0$ are both in radian.

The WFS produces a set of $N$ values $I_0, I_1, ... I_{N-1}$ (usually light intensities, but can also be centroid positions for a SHWFS) which are used to compute $x$ and $y$, the measured values of $x_0$ and $y_0$. Because of photon noise, $x \neq x_0$ and $y \neq y_0$. After correction by the DM, the residual sine wave phase aberration at the spatial frequency considered has an amplitude $A_{res}$ (in radian):
\begin{equation}
A_{res} = \sqrt{(x-x_0)^2+(y-y_0)^2}.
\end{equation}

In order to compute $\Sigma$, the residual phase error in the pupil plane in radians rms, the probability distribution $P(x,y)$ of $x$ and $y$ can be used:
\begin{equation}
\label{equ:Sigma}
\Sigma = \sqrt{\int_{x,y} P(x,y) ((x-x_0)^2+(y-y_0)^2) dxdy}.
\end{equation}

$P(x,y)$ can be written as
\begin{equation}
\label{equ:probprod}
P(x,y) = \prod_{k=0}^{N-1} P_k(x,y),
\end{equation}
where $P_k(x,y)$ is the probability distribution of $x$ and $y$ given by the measurement $I_k$.

I denote ${\sigma_{I_k}}^2$ the variance on the measurement of $I_k$, for which a gaussian probability law is assumed. I also assume that $I_k$ is a linear function of $x$ and $y$ around $(x_0,y_0)$ (this is true in closed-loop AO systems, as $x_0$,$y_0$, $x$ and $y$ are small) :
\begin{equation}
I_k(x,y) = I_k(x_0,y_0) + \frac{d\:I_k}{d\:x} X + \frac{d\:I_k}{d\:y} Y
\end{equation}
where the derivatives are computed in $(x_0,y_0)$, $X=x-x_0$ and $Y=y-y_0$.

In the $(x,y)$ plane, a measurement $I_k$ gives a probability $P_k$ only along one axis (the normalization coefficients of probability distributions are omitted in this section):
\begin{equation}
P_k(x,y) = \exp{-\frac{(k_x\:X+k_y\:Y)^2}{2\:{\sigma_k}^2}}
\end{equation}
where ${k_x}^2+{k_y}^2=1$,
\begin{equation}
\frac{k_x}{k_y} = \frac{d\:I_k}{d\:x}\:\left(\frac{d\:I_k}{d\:y}\right)^{-1},
\end{equation}
and
\begin{equation}
\sigma_k = \sigma_{I_k} \: \left(\sqrt{\left(\frac{d\:I_k}{d\:x}\right)^2+\left(\frac{d\:I_k}{d\:y}\right)^2}\right)^{-1}.
\end{equation}
The vector $(k_x,k_y)$ gives the direction along which $I_k$ is ``sensitive''. For example, if $d\:I_k/d\:x = d\:I_k/d\:y$, $(k_x,k_y)=(1/\sqrt{2},1/\sqrt{2})$, and $P(x,y)$ is only a function of $X+Y$. In this example, the probability is constant along a line $x = cst - y$, as the 2 partial derivatives of $I_k$ cancel when moving along this line. 

From equation \ref{equ:probprod},
\begin{equation}
\label{equ:probxy}
P(x,y) = \exp{-\left( \frac{X^2}{\alpha_1} + \frac{X\:Y}{\alpha_2} + \frac{Y^2}{\alpha_3}\right)}
\end{equation}
where
\begin{equation}
\frac{1}{\alpha_1} = \sum_{k=0}^N \frac{{k_x}^2}{2\:{\sigma_k}^2}
\end{equation}
\begin{equation}
\frac{1}{\alpha_2} = \sum_{k=0}^N \frac{{k_x}\:{k_y}}{{\sigma_k}^2}
\end{equation}
\begin{equation}
\frac{1}{\alpha_3} = \sum_{k=0}^N \frac{{k_y}^2}{2\:{\sigma_k}^2}.
\end{equation}

\begin{figure}[htb]
\includegraphics[scale=0.55]{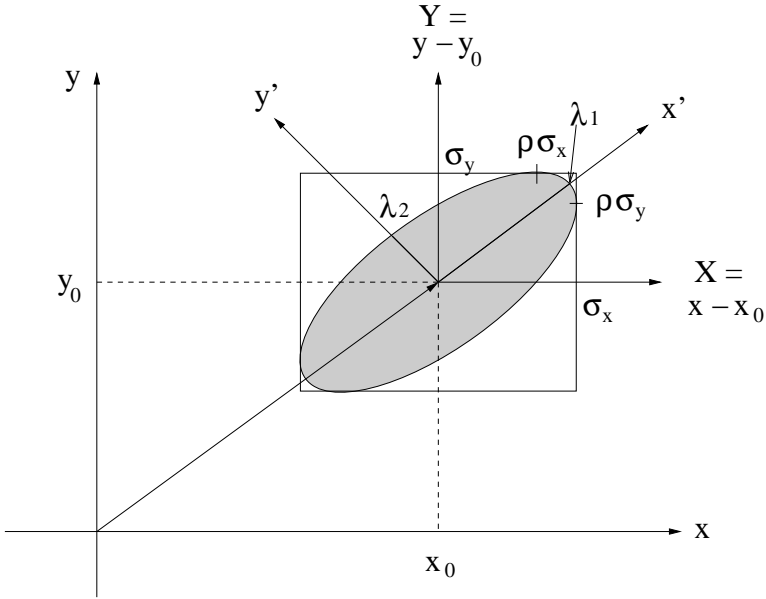}
\caption{\label{fig:pxyell} Graphical representation of the domain $P(x,y)<exp(-1)$ from equation \ref{equ:probxy1}. The domain is circular for $\rho=0$, and elongated for $\rho\neq0$. In this example, $\sigma_x>\sigma_y$ and $\rho>0$. $\lambda_1$ and $\lambda_2$ are the maximal and minimal radii of the ellipse.}
\end{figure}

Equation \ref{equ:probxy} is a 2D normal law :
\begin{eqnarray}
\label{equ:probxy1}
P(x,y) = \exp{\left(\frac{-1}{2(1-\rho^2)} \left[\frac{X^2}{{\sigma_x}^2} - \frac{2 \rho XY}{\sigma_x \sigma_y} + \frac{Y^2}{{\sigma_y}^2}\right]\right)}
\end{eqnarray}
where

\begin{equation}
\rho = \frac{\sqrt{\alpha_1 \alpha_3}}{2\: \alpha_2}
\end{equation}

\begin{equation}
\sigma_x = \sqrt{\frac{\alpha_1}{2\:(1-\rho^2)}}
\end{equation}

\begin{equation}
\sigma_y = \sqrt{\frac{\alpha_3}{2\:(1-\rho^2)}}
\end{equation}

A graphical representation of $P(x,y)$ is shown in figure \ref{fig:pxyell}. In the coordinate system $(x',y')$ which axis are aligned with the long and short axis of the ellipse $P(x,y)=exp(-1)$, 
\begin{equation}
\label{equ:probxy2}
P(x',y') = \exp{\left(-\frac{1}{2}\left(\frac{{x'}^2}{{\lambda_1}^2}+\frac{{y'}^2}{{\lambda_2}^2}\right)\right)}
\end{equation}
where
\begin{equation}
\label{equ:lambda1}
{\lambda_1}^2 = \frac{{\sigma_x}^2+{\sigma_y}^2+\sqrt{({\sigma_x}^2-{\sigma_y}^2)^2+4\:\rho^2\:{\sigma_x}^2\:{\sigma_y}^2}}{2}
\end{equation}
\begin{equation}
\label{equ:lambda2}
{\lambda_2}^2 = \frac{{\sigma_x}^2+{\sigma_y}^2-\sqrt{({\sigma_x}^2-{\sigma_y}^2)^2+4\:\rho^2\:{\sigma_x}^2\:{\sigma_y}^2}}{2}
\end{equation}

From equations \ref{equ:Sigma},\ref{equ:probxy2}, \ref{equ:lambda1} and \ref{equ:lambda2},
\begin{equation}
\Sigma = \sqrt{{\lambda_1}^2 + {\lambda_2}^2}
\end{equation}
\begin{equation}
\Sigma = \sqrt{{\sigma_x}^2+{\sigma_y}^2}
\end{equation}

$\Sigma$ can now be used as a quantitative measure of the ability of a WFS to sense a sine wave phase aberration in the pupil plane. For most WFSs, $\Sigma$ is a function of $\theta$, in which case I consider $\Sigma_{max}$, the maximum value of $\Sigma$ over all values of $\theta$. Since the measurement errors are produced by photon noise,
\begin{equation}
\label{equ:WFS_quality}
\Sigma = \frac{\beta_p}{\sqrt{N_{ph}}}
\end{equation}
where $N_{ph}$ is the total number of photons available for wavefront sensing, and $\beta$ is a function of the WFS. The parameter $\beta_p$ represents the sensitivity of the WFS to photon noise for the spatial frequency considered.

\section{Algebraic representation of WFSs}
\label{app:unif}
In this appendix, I show that any wavefront sensor can be represented as a unitary matrix $U$ and a stochastic matrix $S_c$.

I denote $W(\vec{u})$ the complex amplitude of the incoming wavefront at the position $\vec{u}$ in the pupil.
\begin{equation}
\label{equ:AW}
W(\vec{u}) = \mathscr{A}(\vec{u}) \times e^{i \: \phi(\vec{u})}
\end{equation}
where $\mathscr{A}(\vec{u})$ is the amplitude and $\phi(\vec{u})$ the phase of the wavefront. 
In all WFSs, the incoming wavefront is estimated from measurements of the intensities (square of the modulus of the complex amplitude) obtained by mutual interferences of parts of the incomping wavefronts. One such intensity, $I_k$ can be written as
\begin{equation}
I_k = {|B_k|}^2 = {|\int\limits_{\mathscr{P}} f_k(\vec{u}) W(\vec{u}) \mathrm{d}\vec{u}|}^2 
\end{equation}
where $f_k(\vec{u})$ is a complex function of $\vec{u}$ and 
\begin{equation}
\forall k, {|f_k(\vec{u})|}^2 < 1
\end{equation}
The number of such measurements is $m$. Another constraint is that the incoming light is shared between the different outputs :
\begin{equation}
\vec{r} \in \mathscr{P} \Longrightarrow \sum\limits^{m}_{k=1}{|f_k(\vec{u})|}^2 = 1.
\end{equation}
Finally, the light intensity is conserved between the input (wavefront) and the outputs :
\begin{equation}
\forall \mathscr{A}, \sum\limits^{m}_{k=1} I_k = \int\limits_{\mathscr{P}}{\mathscr{A}(\vec{u})}^2\mathrm{d}\vec{r}
\end{equation}

The continuous incoming wavefront can be approximated by its values on the points of a fine 2D grid : this representation is accurate up to the spatial frequency defined by the spacing of the grid elements. 
In this representation, the incoming wavefront is a vector $A$ (representing $W$ in equation \ref{equ:AW}) which has as many elements $A_k$ as there are evaluation points on the pupil. I denote $n$ the number of such elements. Each function $f_k(\vec{u})$ is represented by a vector $U_k$ of $n$ elements $U_k^l, l=1 \cdots n$.
The measured intensities are a vector $I$ of $m$ elements which is the square of the amplitude of the vector $B$ (elements $B_k$) :
\begin{equation}
I=\left[ \begin{array}l
I_1\\
I_2\\
.\\
.\\
I_m\\
\end{array}\right] = \left| \left[ \begin{array}l
B_1\\
B_2\\
.\\
.\\
B_m\\
\end{array}\right] \right|^2=\left|B\right|^2
\end{equation}
with 
\begin{equation}
B = UA =\left[ \begin{array}{llll}
U_1^1&U_1^2&\cdots&U_1^m\\
U_2^1&U_2^2&\cdots&U_2^m\\
\cdots&\cdots&\cdots&\cdots\\
U_n^1&U_n^2&\cdots&U_n^m\\
\end{array}\right]
\left[ \begin{array}l
A_1\\
A_2\\
.\\
.\\
A_n\\
\end{array}\right]
\end{equation}
The conservation of total light intensity requires
\begin{equation}
\forall A, \|UA\|=\|A\|.
\end{equation}
Setting $m=n$, $U$ is therefore a unitary matrix.
In all current WFSs, $m$ is in fact infinite and the measured quantity is not directly $I$, but a set of intensities obtained by redistributing the values of $I$ among a smaller number of variables, which can be represented by a stochastic matrix $S_c$ preserving the total flux. For example, in a SHWFS, the intensity measured by a pixel is in fact the integral of the intensity accross the pixel.


\begin{thebibliography}{}
\bibitem[Aime \& Soummer(2004)]{aime04} Aime, C., Soummer, R.  2004, \apjl, 612, L85
\bibitem[Angel(1994)]{ange94} Angel, J.R.P. 1994, \nat, 368, 203 
\bibitem[Angel(2003)]{ange03} Angel, J.R.P.  2003, ASP Conf. Ser, 294, 543-556, ``Scientific Frontiers in Research on Extrasolar Planets'', eds. S. Seager and D. Deming, Washington D.C. 2003
\bibitem[Aristidi et al.(2005)]{aris05} Aristidi, E., Agabi, K., Azouit, M., Fossat, E., Vernin, J., Travouillon, T., Lawrence, J. S., Meyer, C., Storey, J. W. V., Halter, B., Roth, W. L., Walden, V.  2005, \aa, 430, 739
\bibitem[Baudoz et al.(2000)]{baud00} Baudoz, P., Rabbia, Y., Gay, J. 2000, \aaps, 141, 319
\bibitem[Bloemhof et al.(2001)]{bloe01} Bloemhof, E.E., Dekany, R.G., Troy, M., Oppenheimer, B.R.  2001, \apjl, 558, L71
\bibitem[Bloemhof(2003)]{bloe03} Bloemhof, E.E.  2003, \apjl, 582, L59
\bibitem[Bloemhof \& Wallace(2003)]{bloe03a} Bloemhof, E.E., Wallace, J.K., \procspie, 5169, 309
\bibitem[Boccaletti et al.(1998)]{bocc98} Boccaletti, A., Moutou, C., Labeyrie, A., Kohler, D., Vakili, F.  1998, \aaps, 133, 395
\bibitem[Boccaletti et al.(2003)]{bocc03} Boccaletti, A., Augereau, J.-C., Marchis, F., Hahn, J. 2003, \apj, 585, 494
\bibitem[Codona \& Angel(2004)]{codo04} Conona, J., \& Angel, R.   2004, \apjl, 604, L117
\bibitem[Edlen(1966)]{edle66} Edlen, B.  1966, {\it Metrologia}, 2, 71
\bibitem[ESO(2004)]{eso04} European Southern Observatory, http://www.eso.org/gen-fac/pubs/astclim/paranal/seeing/adaptive-optics/
\bibitem[Faucherre et al.(1989)]{fauc89} Faucherre, M. , Merkle, F. , Vakili, F.  1989, \procspie, 1130, 138
\bibitem[Flicker \& Rigaut(2002)]{flic02} Flicker, R.C., Rigaut, F.J.  2002, \pasp, 114, 1006
\bibitem[Gerchberg \&  Saxton(1972)]{ger72} Gerchberg, R.W., Saxton, W.O.  1972, Optik, 35, 237
\bibitem[Gilmozzi(2004)]{gilm04} Gilmozzi, R., 2004, \procspie, 5489, 1
\bibitem[Guyon(2004)]{guyo04} Guyon, O.  2004 \apj, 615, 562
\bibitem[Guyon et al.(2005)]{guyo05} Guyon, O., Pluzhnik, E.A., Galicher, R., Martinache, F., Ridgway, S.T, Woodruff R.A.  2005, \apj, 622
\bibitem[Hardy(1998)]{hard98} Hardy, J.  1998, ``Adaptive Optics for Astronomical Telescopes'', Oxford University Press, p. 398.
\bibitem[Hawarden et al.(2003)]{hawa03} Hawarden, T.G., Dravins, D., Gilmore, G.F., Gilmozzi, R., Hainaut, O., Kuijken, K., Leibindgut, B., Merrifield, M., Queloz, D., Wyse, R.  2003, \procspie, 4840, 299
\bibitem[Kasdin et al.(2003)]{kasd03} Kasdin, N.J., Vanderbei, R.J., Spergel, D.N., Littman, M.G. 2003, \apj, 582, 1147
\bibitem[Kuchner \& Spergel(2003)]{kuch03} Kuchner, M.J., Spergel, D.N. 2003, 594, 617
\bibitem[Leger et al.(1997)]{lege97} Leger, J.R., Schuler, J., Morphis, N., Knowlden, R.  1977, Applied Opt., 36, 4692L
\bibitem[Pickles(1998)]{pick98} Pickles, A.  1998, \pasp, 110, 863
\bibitem[Poyneer \& Macintosh(2004)]{poyn04} Poyneer, L.A., Macintosh, B.  2004, J. Opt. Soc. Am., 21, 810
\bibitem[Lawrence et al.(2004)]{lawr04} Lawrence, J.S., Ashley, M.C.B., Tokovinin, A., Travouillon, T. 2004, \nat, 431, 278
\bibitem[Malbet et al.(1995)]{malb95} Malbet, F., Yu, J.W., \& Shao, M.  1995, \pasp, 107, 386
\bibitem[Marois et al.(2000)]{maro00} Marois, C., Doyon, R., Racine, R., \& Nadeau, D.  2000, \pasp, 112, 91
\bibitem[Racine \& Ellerbroek(1995)]{raci95} Racine, R., Ellerbroek, B.L.  1995, \procspie,  2534, 248
\bibitem[Racine et al.(1999)]{raci99} Racine, R., Walker, G.A.H., Nadeau, D., Doyon, R., \& Marois, C.  1999, \pasp, 111, 587
\bibitem[Ragazzoni(1996)]{raga96} Ragazzoni, R.  1996 J. Mod. Opt., 43, 289
\bibitem[Roddier et al.(1980)]{rodd80} Roddier, C., Roddier, F., Martin, F., Baranne, A., Brun, R.  1980, Journ. of Optics, 11, 149
\bibitem[Roddier(1988)]{rodd88} Roddier, F.  1988, Appl. Opt., 27, 1223
\bibitem[Roddier et al.(1991)]{rodd91} Roddier, F., Northcott, M., Graves, J.E.  1991, \pasp, 103, 131 
\bibitem[Roddier(1997)]{rodd97} Roddier, F., Roddier, C. 1997, \pasp, 109, 815
\bibitem[Rouan et al.(2000)]{roua00} Rouan, D., Riaud, P., Boccaletti, A., Clénet, Y., Labeyrie, A. 2000, \pasp, 112, 1479
\bibitem[Roth et al.(2001)]{roth01} Roth, K.C., Guyon, O., Chun, M., Jensen, J.B., Jorgensen, I., Rigaut, F., Walther, D.M. 2001, \baas, 33, 785
\bibitem[Soummer et al.(2003)]{soum03} Soummer, R., Aime, C., Falloon, P.E. 2003, \aap, 397, 1161
\bibitem[Soummer et al.(2003a)]{soum03a} Soummer, R., Dohlen, K., Aime, C. 2003, \aap, 403, 369
\bibitem[Subaru Telescope(2004)]{naoj04} Subaru Telescope, National Astronomical Observatory of Japan, http://www.naoj.org/Observing/Telescope/Image/seeing.html
\bibitem[Talbot(1836)]{talb36} Talbot, W.H.F.  1836, ``Facts relating to optical science, No. IV'',  Philos. Mag. 9, 401
\bibitem[Takato(2005)]{taka05} Takato, N. 2005, private communication
\bibitem[Tokovinin et al.(2005)]{toko05} Tokovinin A., Vernin J., Ziad A., Chun M. 2005, \pasp, in press
\bibitem[V\'erinaud(2004)]{veri04} V\'erinaud, C.  2004, Opt. comm., 223, 27
\bibitem[V\'erinaud et al.(2005)]{veri05} V\'erinaud, C., Le Louarn, M., Korkiakoski, V., Carbillet, M.  2005, \mnras, 357, 26
\bibitem[Wynne(1979)]{wynn79} Wynne, C.G.  1979, Opt. Comm., 28, 21
\bibitem[Yang \& Kostinski(2004)]{yang04} Yang, W., Kostinski, A.B. 2004, \apj, 605, 892
\bibitem[Zernike(1934)]{zern34} Zernike, F.  1934, \mnras, 94, 377
\end{thebibliography}
\end{document}